\documentclass[aps,prx,showpacs,twocolumn,reprint,superscriptaddress]{revtex4-2}

\usepackage[dvips]{graphicx}
\usepackage{amsmath,amssymb,amsthm,mathrsfs,amsfonts,dsfont,mathtools}
\usepackage{epsfig}
\usepackage{braket}
\usepackage{hyperref}
\usepackage{bm}
\usepackage{enumerate}
\usepackage{color}
\usepackage{graphicx}
\usepackage{pgf}
\usepackage{pgfplots}
\pgfplotsset{compat=1.18}
\usepackage{tikz}
\usepackage{enumitem}
\usepackage[capitalize]{cleveref}
\usepackage[normalem]{ulem}
\usepackage{physics}
\usetikzlibrary{quantikz2}
\usepackage[caption=false]{subfig}

\newcommand{\dde}{DDE}

\usepackage{amsthm}

\newtheorem{proposition}{Proposition}
\newtheorem{theorem}{Theorem}
\newtheorem{lemma}{Lemma}

\crefname{theorem}{Theorem}{Theorems}
\theoremstyle{remark}

\newtheoremstyle{colored}
  {3pt}   
  {3pt}   
  {\itshape}  
  {}     
  {\color{blue}\bfseries} 
  {.}    
  { }    
  {}     

\newcommand{\ravg}{\bar\rho}

\usepackage{xcolor}

\begin{document}	
	
	\title{Exponential distillation of dominant eigenproperties}

        \author{Bence Bak\'o}
        \affiliation{Mathematical Institute, University of Oxford, Woodstock Road, Oxford OX2 6GG, United Kingdom}
        \affiliation{HUN-REN Wigner Research Centre for Physics, Konkoly–Thege Mikl\'os \'ut 29-33, Budapest, H-1525, Hungary}
        \affiliation{Faculty of Informatics, E\"otv\"os Lor\'and University, P\'azm\'any P\'eter s\'et\'any 1/C, Budapest, H-1117, Hungary}

        \author{Tenzan Araki}
        \affiliation{Mathematical Institute, University of Oxford, Woodstock Road, Oxford OX2 6GG, United Kingdom}
        \affiliation{Department of Physics, Clarendon Laboratory, University of Oxford, Parks Road, Oxford OX1 3PU, United Kingdom}
 
	\author{B\'alint Koczor}
	\email{balint.koczor@maths.ox.ac.uk}
        \affiliation{Mathematical Institute, University of Oxford, Woodstock Road, Oxford OX2 6GG, United Kingdom}
	
	\begin{abstract}
Estimating observable expectation values in eigenstates of quantum systems has a broad range of applications and is an area where early fault-tolerant quantum computers may provide practical quantum advantage. We develop a hybrid quantum-classical algorithm that enables the estimation of an arbitrary observable expectation value in an eigenstate, given an initial state is supplied that has dominant overlap with the targeted eigenstate -- but may overlap with any other eigenstates. Our approach builds on and is conceptually similar to purification-based error mitigation techniques; however, it achieves exponential suppression of algorithmic errors using only a single copy of the quantum state. The key innovation is that random time evolution is applied in the quantum computer to create an average mixed quantum state, which is then virtually purified with exponential efficacy. We prove rigorous performance guarantees and conclude that the complexity of our approach depends directly on the energy gap in the problem Hamiltonian and remarkably, can be compared to phase estimation combined with amplitude estimation in terms of its scaling with respect to a target precision. We demonstrate in a broad range of numerical simulations the applicability of our framework in near-term and early fault-tolerant settings. Furthermore, we demonstrate in a 100-qubit example that direct classical simulation of our approach enables the prediction of ground and excited state properties of quantum systems using tensor-network techniques, which we recognize as a quantum-inspired classical approach.
	\end{abstract}

	\maketitle

\section{Introduction}\label{sec:intro}
Estimating eigenstate properties of quantum many-body systems is a central task in quantum chemistry, material science, high-energy
physics and beyond~\cite{cao2019quantum,mcardle2020quantum,bauer2020quantum,motta2022emerging}.
While quantum computers may offer an exponential advantage over classical techniques when a good initial state is known~\cite{low2016methodology,low2017optimal,lin2020near},
significant challenges remain~\cite{kempe2005complexity,lee2023evaluating}.
For example, typical fault-tolerant quantum algorithms require deep circuits beyond the capabilities of early fault-tolerant machines, expected to emerge in the near term~\cite{babbush2018encoding,gidney2021how}. 
In contrast, variational quantum algorithms are suitable for near-term quantum devices
as heuristic alternatives to investigate ground~\cite{cerezo2021variational,Endo2021,bharti2022noisy,mcardle2019variational,PhysRevA.106.062416} and excited state properties~\cite{higgott2019variational,Boyd_2022,hwang2024preparing}; 
however, they may suffer from large shot-number requirements, from getting stuck in local minima, and
from barren plateaus~\cite{larocca2025barren, anschuetz2022quantum,tikku2022circuit}.

To bridge the gap between variational techniques and fully fault-tolerant quantum algorithms,
a range of promising hybrid algorithms have been developed that rely on classically post-processing data obtained
from randomly sampled quantum circuits with moderate quantum resource requirements.
These include techniques based on classical shadows~\cite{boyd2025high, Boyd_2022,hwang2024preparing,jnane2024quantum},
statistical phase estimation~\cite{lin2022heisenberg, wang2023quantum, wang2025efficient,kiss2025early,PRXQuantum.6.010352}
and on Monte Carlo (MC) methods \cite{mazzola2024quantum, huo2023error, zeng2022universal, sun2024high, yang2021accelerated}.

\begin{figure}
    \centering
    \includegraphics[width=.48\textwidth]{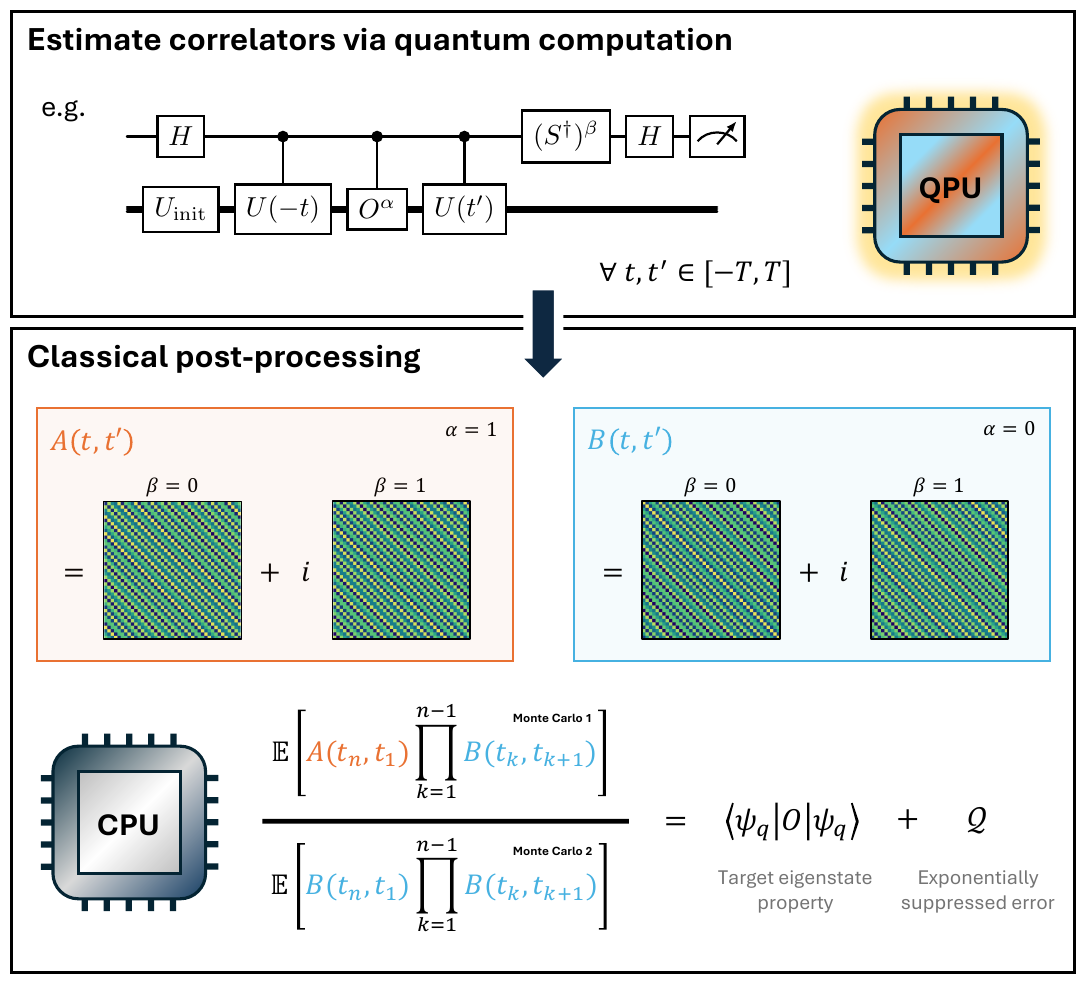}
    \caption{DDE proceeds by estimating two-time correlators $A(t,t') = \bra{\psi(t)} O \ket{\psi(t')}$ and $B(t,t') = \bra{\psi(t)} \ket{\psi(t')}$ using Hadamard test circuits over a 2D time grid in $t$ and $t'$. Then, Monte Carlo sampling is used to evaluate high-dimensional integrals, which lead to the estimation of an observable expectation value in a target eigenstate of a problem Hamiltonian. \cref{theo:joint-bound} guarantees that the error $\mathcal{Q}$ of the final estimate is suppressed exponentially with respect to chosen hyperparameters.}
    \label{fig:algo}
\end{figure}

In this work, we present a hybrid quantum-classical algorithm (summarized in \cref{fig:algo})
that we call distillation of dominant eigenproperties (\dde).  Given an input state that may have overlap with a large number of eigenstates, but has dominant overlap with a targeted eigenstate, our approach enables the estimation of observable expected values in the targeted eigenstate -- therefore, generalizing statistical phase estimation type approaches that are limited to eigenenergy estimation~\cite{lin2022heisenberg,wang2023quantum, kiss2025early,PRXQuantum.6.010352}.
Our approach is inspired by purification-based quantum error mitigation (QEM) techniques \cite{koczor2021exponential, huggins2021virtual}
and we show that \dde\ exponentially suppresses contributions from non-dominant eigenstates, but without having to prepare multiple copies. It achieves a fundamentally reduced circuit depth at the cost of a reasonably increased number of quantum circuit repetitions, and is therefore ideally positioned for early fault-tolerant devices. While related works can
estimate expected values using similar quantum resource requirements~\cite{huo2023error,zeng2022universal,sun2024high,yang2021accelerated},
as we detail in Appendix~\ref{appendix:related}, our approach is not based on imaginary time evolution; therefore, it is not limited to the ground state and requires no explicit knowledge of the relevant eigenenergies of the problem Hamiltonian.
Furthermore, while \dde\ is more generally applicable than statistical phase estimation algorithms, it achieves optimally shallow circuit depths $\tilde{\mathcal{O}}(\Delta^{-1})$, comparable to state-of-the art energy estimation approaches~\cite{wang2023quantum, wang2025efficient}.
However, we detail in Appendix~\ref{appendix:related} that estimating arbitrary expectation values in eigenstates has an increased complexity compared to estimating the energy alone. Remarkably, our approach for estimating arbitrary expectation values achieves a scaling (with respect to the precision) that is comparable to that of fault-tolerant phase estimation combined with amplitude estimation. 

DDE relies on estimating two-time correlations that can be computed using a quantum device and subsequently processed classically.
We need only assume the availability of a quantum state with dominant overlap with the targeted eigenstate
and a finite energy gap between the target eigenstate and any other eigenstate in the support of the initial state, which is a common assumption across many
quantum simulation techniques~\cite{low2025fast, berry2025rapid, morchen2024classification}. Obtaining a good initial state is a challenging task in general, based on complexity theoretic results~\cite{bookatz2012qma}. However, very successful classical approximation methods have been developed for a broad range of important practical problems in quantum chemistry, solid-state physics, and material science, such as matrix product states, Hartree-Fock, and post-Hartree-Fock techniques. The community has identified a range of problems where 
no known classical approximations are sufficiently accurate to be useful, but can still be used to provide a reasonably good initial state
for quantum computation~\cite{zimboras2025myths, babbush2025grand}. The primary example here is FeMoco, which is a key target for substantial early practical quantum advantage~\cite{low2025fast, berry2025rapid} -- classical tensor-network approximations have been used to prepare an initial state with above 90\% fidelity, but quantum computing is necessary to improve upon this accuracy in order to be useful in practical applications.

Our general framework is compatible with a broad range of initial state preparation and time evolution techniques. We demonstrate in explicit numerical simulations that the initial state can be successfully prepared using Hartree-Fock or matrix product state (MPS)~\cite{low2025fast, berry2025rapid, morchen2024classification}
approximations, or through quantum heuristics such as the variational quantum eigensolver (VQE)~\cite{peruzzo2014variational}.
We also demonstrate that time evolution can be implemented using near-term methods, such as variational quantum simulation \cite{yuan2019theory}, or early fault-tolerant techniques, such as product formulas~\cite{trotter1959product,suzuki1976generalized}.

The rest of the paper is organized as follows: In \cref{sec:preliminaries}, we first present the theoretical basis of DDE and our rigorous performance guarantees. The main results are then presented in \cref{sec:results}, in which DDE is formally introduced, and its implementation is detailed. We demonstrate in \cref{sec:demo} the utility of DDE through numerical simulations, when combined with existing techniques in both ideal and more realistic hardware implementations. In \cref{sec:classical}, we detail a novel classical simulation technique for estimating
ground and excited state energies in a 100-qubit spin system using tensor-network classical simulation of our DDE approach.
We finally conclude in \cref{sec:discussion}.

\section{Theoretical Guarantees}
\label{sec:preliminaries}

\begin{figure*}[t]
\centering
\includegraphics[width=0.8\textwidth]{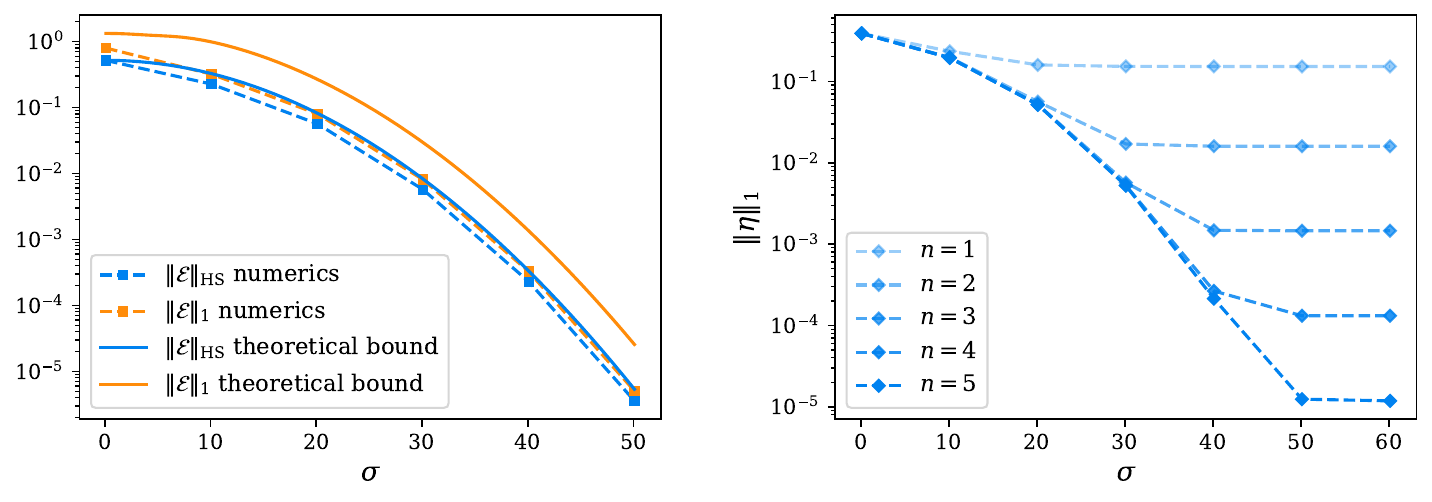}

  \caption{\textbf{Numerically verifying our error bounds.}
  Using an initial state with $p_q \approx 0.85$ and a time grid defined on $[-5\sigma, 5\sigma]$ with step size $dt = 1$, we apply temporal averaging in a $12$-qubit random-field Heisenberg model simulation. Error contributions are shown against the standard deviation $\sigma$. (left) Verifying \cref{lemma:time-evolution}: numerically computed error $\mathcal{E} =\ravg-\rho$ norms (dashed lines) are indeed below their corresponding analytical upper bounds (solid lines), and decrease super-exponentially as we increase the standard deviation of the normal distribution. (right) Numerically demonstrating \cref{theo:joint-bound}, the error scaling, as measured in the trace norm of $\eta = \bar\rho ^n / \mathrm{tr}[\bar\rho ^n] - \ketbra{\psi_q}$, is shown for an increasing number of copies, where $\ket{\psi_q}$ denotes the target eigenstate. As stated in \cref{theo:joint-bound}, this error norm is a sum of two terms: the first term is suppressed exponentially as we increase the number of copies $n$, whereas the second term is  suppressed super-exponentially 
  as we increase $\sigma$, as illustrated above.
}
  \label{fig:theory}
\end{figure*}

The hybrid \dde\ algorithm proposed in this work is formally related to, and draws inspiration from,
the QEM technique virtual distillation (VD) \cite{koczor2021exponential, huggins2021virtual}.
Given $n$ copies of a mixed state that correspond to the noisy execution of a quantum circuit, VD effectively purifies it to exponentially suppress erroneous contributions in expected values. The crucial difference is that the present DDE algorithm deliberately creates a mixed state out of a set of randomly prepared pure states. In the following, we derive new results to show that sampling random evolution times enables the preparation of a mixed quantum state that is nearly diagonal in the energy eigenbasis of a pre-specified problem Hamiltonian. Then, virtually purifying this state as in Refs.~\cite{koczor2021exponential,huggins2021virtual} allows the estimation of eigenstate properties with exponential convergence guarantees.

\subsection{Random time evolution}
Consider a Hamiltonian expressed in its spectral decomposition as $\mathcal{H} = \sum_k E_k \ketbra{\psi_k}$, with eigenenergies
$E_k$ and energy eigenstates $|\psi_k\rangle$.
An arbitrary initial state can be expressed as
$\ket{\psi(0)} = \sum_k c_k \ket{\psi_k}$.
Time evolving this initial state under $\mathcal{H}$ leads to
$\ket{\psi(t)} = \sum_k c_k e^{-i t E_k} \ket{\psi_k}$,
and we define the corresponding density matrix as $\rho(t) = \ketbra{\psi(t)}$.

We also define the state $\rho := \sum_k p_k \ketbra{\psi_k}$, which is diagonal in the energy eigenbasis with eigenvalues related to the initial state amplitudes as $p_k = |c_k|^2$. The following lemma guarantees that if we randomly sample time-evolved states $\rho(t)$  such that $t$ is chosen from an appropriate normal distribution, then the average state approximates $\rho$.

\begin{lemma}
Averaging the time-evolved states $\rho(t) = e^{-i t \mathcal{H}} \ketbra{\psi(0)} e^{i t \mathcal{H}}$ over evolution times $t$ that are randomly sampled according to a normal distribution $\mathcal{N}(0,\sigma)$ leads to a mixed state $\ravg$ given by
\begin{equation}\label{eq:t_integral}
	\ravg = \mathds{E}_{t}[\rho(t)] = \int_{-\infty}^{\infty} G(t) \rho(t) \, \mathrm{d}t = \rho + \mathcal{E},
\end{equation}
where $G(t) = (\sqrt{2 \pi } \sigma )^{-1}  e^{-\frac{t^2}{2 \sigma ^2}}$ is the probability density function (PDF) of the normal distribution, and $\rho = \sum_k p_k \ketbra{\psi_k}$ is a diagonal state with eigenvalues $p_k = |c_k|^2$ determined by the initial state $\ket{\psi(0)} = \sum_k c_k \ket{\psi_k}$.
The error term $\mathcal{E}$ decreases super-exponentially as we increase the standard deviation as $\lVert \mathcal{E} \rVert_{\mathrm{HS}}  \leq e^{-\frac{\Delta^2 \sigma^2}{2}}\sqrt{ \left(1 - e^{-H_2(\bm{p})} \right)}$,
where
\begin{equation}
\Delta = \min_{j,k: \,|c_j|,|c_k|> 0}\{ |E_j-E_k| \}
\end{equation}
 is the smallest energy gap in the support of the initial state. Furthermore, $H_2(\bm{p})$ is the second R\'enyi entropy over $\bm{p} = (p_1,p_2,p_3,...)$ (see \cref{eq:renyi}), and $||\cdot||_{\mathrm{HS}}$ is the usual Hilbert-Schmidt norm. 
The trace distance is similarly bounded as $\lVert \mathcal{E} \rVert_{1} =  \lVert \ravg - \rho \rVert_{1} \leq e^{-\frac{\Delta^2 \sigma^2}{2}} \lVert \bm c \rVert_1$, where 
$\lVert \bm c \rVert_1$ is the $\ell_1$ norm of the coefficient vector $\bm c = (c_1,c_2,c_3,...)$ in the eigenbasis of the residual operator $\mathcal{E}$.
\label{lemma:time-evolution}
\end{lemma}

Please refer to Appendix~\ref{appendix:proofs} for a detailed proof of the Hilbert-Schmidt and the trace distance bounds. The Gaussian window function in the integral enables us to truncate the time grid to $[-T,T]$ for a sufficiently large $T$, e.g., $T=5 \sigma$.

In \cref{fig:theory} (left), we numerically verify the above error bounds by simulating a $12$-qubit random-field
Heisenberg model with periodic boundary conditions -- we defer details of the simulation to Appendix~\ref{appendix:spinchain}.
We plot both the Hilbert-Schmidt distance $\lVert \mathcal{E} \rVert_{\mathrm{HS}} $ and the trace distance $\lVert \mathcal{E} \rVert_{1} $
(dashed lines) for an increasing standard deviation $\sigma$,
and confirm that they are indeed below their theoretical upper bounds (solid lines).

Most prominent quantum simulation algorithms, such as quantum phase estimation, assume that an initial state that has a dominant overlap with a target eigenstate $\ket{\psi_q}$ can be prepared efficiently. In these typical applications, the initial state is low in energy and, therefore, only has support on the few lowest-lying eigenstates; thus $\Delta$ in the error bounds in \cref{lemma:time-evolution} corresponds to the smallest gap among the low-lying eigenstates. 
In complete generality, the error term $\mathcal{E}$ can be suppressed efficiently as long as the initial state only overlaps with eigenstates that are well separated in energy. In such instances, $\rho$ is of low rank, and we obtain a tighter general bound on the trace distance as $\lVert \mathcal{E} \rVert_{1} \leq 2\sqrt{\mathrm{rank}(\bar{\rho})}\lVert \mathcal{E} \rVert_{\mathrm{HS}} \leq 2\sqrt{ \mathrm{rank}(\rho)}\lVert \mathcal{E} \rVert_{\mathrm{HS}}$. Since $\mathrm{rank}(\rho)$ is equal to the support of $\bm p$, this prefactor expresses how many eigenstates the initial state overlaps with.

Furthermore, the approach becomes more efficient as we increase the overlap of the initial with the target state.
Specifically, the Hilbert-Schmidt distance bound in \cref{lemma:time-evolution} depends on the R\'enyi entropy
\begin{equation}
     H_a(\bm{p}) \coloneqq \frac{1}{1-a}\ln{\left[ \sum_{k} p_k^a\right]}
     \label{eq:renyi}
\end{equation}
of order $a=2$, which can be upper bounded as
\begin{equation}
H_2(\bm{p}) \leq 2H_{\infty}(\bm{p}) = -2\ln{[\max\limits_{k} p_k]}.
\end{equation}
Substituting the above bound back to \cref{lemma:time-evolution} confirms that, indeed, $\lVert \mathcal{E} \rVert_{\mathrm{HS}}$ vanishes in the limit where the initial state is exactly an eigenstate of the Hamiltonian
via $H_2(\bm{p}) \rightarrow 0$.
In the rest of this work, we will assume that the initial state has a dominant overlap with a target eigenstate $\ket{\psi_q}$, and we will demonstrate a range of practical approaches for the preparation of such initial states in \cref{sec:demo}. Our error bound, then also scales with this dominant overlap $p_q = |\bra{\psi_q}\ket{\psi(0)}|^2$ as $H_2(\bm{p}) \leq -2\ln{[ p_q]}.$

\subsection{Applying virtual distillation}
Refs.~\cite{koczor2021exponential,huggins2021virtual} presented an approach for estimating non-linear functionals of the density matrix of the form $\tr[\rho^n O]$, where $O$ is a Hermitian observable.
We now use our result from Lemma~\ref{lemma:time-evolution} to prepare an average mixed state that has dominant overlap with
one specific eigenstate $\ket{\psi_q}$ and apply Theorem 2 of Ref.~\cite{koczor2021exponential}, which we adapt for the present context as the following lemma.
\begin{lemma}[Theorem 2 of Ref.~\cite{koczor2021exponential}]
\label{lemma:esd}
Consider the mixed state $\rho = \sum_k p_k \ketbra{\psi_k}$ and assume that it has a dominant overlap $p_q$ with a target eigenstate $\ket{\psi_q}$. Applying results of Refs.~\cite{koczor2021exponential,huggins2021virtual} for estimating non-linear functionals $\tr[\rho^n O]$ for a Hermitian observable $O$, the expectation value $\bra{\psi_q}O\ket{\psi_q}$ can be estimated as
\begin{equation}\label{eq:vd}
	\frac{\tr[\rho^n O]}{ \tr[\rho^n] } = \bra{\psi_q} O \ket{\psi_q} + \mathcal{A},
\end{equation}
where the error term $\mathcal{A}$ is exponentially suppressed via the explicit expression from Ref.~\cite{koczor2021exponential} as $|\mathcal{A}| \leq 2 (p_q^{-1}-1)^n \exp{-(n-1)H_n(\bm p')}$, where $p'_k = p_k / \sum_{i \neq q} p_i$ for $k\neq q$. 
\end{lemma}

Using our results from \cref{lemma:time-evolution} and the results of Refs.~\cite{koczor2021exponential,huggins2021virtual}, as summarized above in \cref{lemma:esd}, we can apply VD to the time-averaged state $\ravg$ to estimate the target eigenstate property $\bra{\psi_q} O \ket{\psi_q}$ with provably exponential convergence guarantees. The joint error bound is summarized in the following theorem.

\begin{theorem}\label{theo:joint-bound}
    Given $n$ copies of the time-averaged state $\ravg$ from \cref{lemma:time-evolution}, one can apply virtual distillation as in \cref{lemma:esd} to estimate the expectation value of the Hermitian observable $O$ as
    \begin{equation}
        \frac{\mathrm{tr}[\ravg^nO]}{\mathrm{tr}[\ravg^n]} = \bra{\psi_q} O \ket{\psi_q} + \mathcal{Q},
    \end{equation}
    where the error $\mathcal{Q}$ can be bounded as
    \begin{equation}\label{eq:error_terms}
        |\mathcal{Q}| \leq |\mathcal{A}| + 4 n p_q^{-1} \| O \|_{\infty} \| \mathcal{E} \|_1 + \mathcal{O}(\|\mathcal{E}\|_1^2).
    \end{equation}
    Substituting our bounds on $\Vert\mathcal{E}\rVert_1$ and $|\mathcal{A}|$ from \cref{lemma:time-evolution} and \ref{lemma:esd}, respectively, we obtain
       \begin{align*}
    	|\mathcal{Q}| \leq\; &2 (p_q^{-1}-1)^n \exp{-(n-1)H_n(\bm p')}\\
    	 &+ 4 n p_q^{-1} 
    	  e^{-\frac{\Delta^2 \sigma^2}{2}} \lVert \bm c \rVert_1 \| O \|_{\infty}  
    	  + \mathcal{O}(\|\mathcal{E}\|_1^2),
    \end{align*}
    which shows explicit dependence on the absolute largest observable eigenvalue $\| O \|_{\infty}$,
    the probability distribution $p'_k = p_k / \sum_{i \neq q} p_i$ for $k\neq q$, and $\lVert \bm c \rVert_1$, the $\ell_1$ norm of the coefficient vector $\bm c = (c_1,c_2,c_3,...)$ from \cref{lemma:time-evolution}.
    Finally, suppressing errors to a desired level of precision $|\mathcal{Q}|$ requires the following number of copies and time evolution window depth (quantum resources):
    \begin{equation}\label{eq:thrm1_resource}
        n = \mathcal{O}( \ln [|\mathcal{Q}|^{-1}])
        \quad and \quad
        \sigma = \mathcal{O}( \Delta^{-1} \sqrt{\ln [|\mathcal{Q}|^{-1}n]} ),
    \end{equation}
    ignoring logarithmic factors.
\end{theorem}

Please refer to Appendix~\ref{appendix:proofs} for a proof. Indeed, our approach matches the $\tilde{\mathcal{O}}(\Delta^{-1})$ circuit depth scaling of state-of-the-art phase estimation protocols~\cite{lin2022heisenberg, wang2023quantum}, which we further detail in Appendix~\ref{appendix:related}.
The first term in the above error bound can be suppressed exponentially by increasing $n$. As we discuss in the next section, the suppression of this first term does not require additional quantum resources, as the number of (virtual) copies only affects classical post-processing in our approach. Therefore, only the suppression of the second error term (along with the higher order terms) requires an increase in quantum resources through increasing the standard deviation $\sigma$, and consequently the time of the quantum simulation.

We confirm the above bounds in \cref{fig:theory} (right) by plotting the trace distance $\| \ravg^n / \mathrm{tr}[\ravg^n] - \ketbra{\psi_q}\|_1$, which generally upper bounds the left-hand side of  \cref{eq:error_terms} independently of the observable via
\begin{equation*}
	\left|\frac{\mathrm{tr}[\ravg^nO]}{\mathrm{tr}[\ravg^n]} - \bra{\psi_q} O \ket{\psi_q}\right|
	\leq 2 \lVert O \rVert_{\infty} \left\lVert \frac{\ravg^n}{\mathrm{tr}[\ravg^n]} - \ketbra{\psi_q}\right\rVert_{1}.
\end{equation*}
Refer to Ref.~\cite{koczor2021dominant} for a detailed proof of the above bound. 
Indeed, \cref{fig:theory} (right) nicely illustrates that 
our error bound in \cref{theo:joint-bound} is a sum of two terms: 
The first term $|\mathcal{A}|$ is suppressed exponentially as we increase $n$, whereas
the second term is increased linearly in $n$ but gets super-exponentially suppressed as
we increase the Gaussian window $\sigma$.

\section{Distillation of dominant eigenproperties}
\label{sec:results}

\cref{theo:joint-bound} allows us to apply VD to randomly time-evolved initial states
in order to estimate eigenstate properties with exponential convergence guarantees as we increase the number of copies. In this section, we describe our approach for achieving this convergence through the DDE algorithm. In particular, we show that virtual purification of randomly time-evolved states can be performed classically via MC integration, without requiring multiple copies of the quantum state. This way, our approach separates into a quantum computation step and a classical post-processing step. We also provide theoretical guarantees for the variance of the MC estimator.

\subsection{Suppressing non-dominant contributions by estimating non-linear functionals\label{subsection: suppressing}}
\begin{figure}
    \centering
    \includegraphics[width=.35\textwidth]{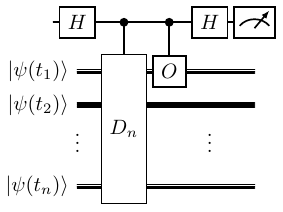}
    \caption{Quantum circuit that enables estimating eigenstate properties, directly following \cref{lemma:time-evolution} and \cref{lemma:esd}. The initial state $\ket{\psi(0)}$ is time evolved for a randomly sampled duration in each of the registers before a derangement (denoted $D_n$) and the observable $O$ are applied, both conditioned on the state of the ancilla qubit. While this implementation requires $n$ quantum registers, DDE only requires $1$ register for any $n$ by using either the circuits shown in \cref{fig:algo} or \cref{fig: QC}.
        }
    \label{fig:vd_circuit}
\end{figure}

Given $n$ copies of the time-averaged state $\ravg$ from \cref{lemma:time-evolution},
we can estimate the required non-linear functionals in \cref{lemma:esd} via the circuit in \cref{fig:vd_circuit}. 
Specifically, the $n$ input states are randomly time-evolved pure states $\ket{\psi(t_i)}$ for $i \in \{1,2,\ldots,n\}$, $D_n$ is a cyclic permutation or derangement operator~\cite{koczor2021exponential}, and $O$ is the observable of interest. The controlled $O$ operation is implemented either by applying individual
Pauli terms in the decomposition of $O$ or via a block encoding using the LCU approach~\cite{childs2012hamiltonian}. Estimating the ancilla probability then allows us to estimate the non-linear functionals $\tr[ \ravg^n O]$ required in \cref{lemma:esd}, with its complexity summarized in \cref{lemma:complexity_vd}.

\begin{lemma}
\label{lemma:complexity_vd}
Sampling the ancilla qubit in \cref{fig:vd_circuit} allows us to estimate $\bra{\psi_q}O\ket{\psi_q}$ to statistical uncertainty $\epsilon$ and bias $Q$ using time evolutions of depth $\sigma$, and number of copies $n$ from \cref{eq:thrm1_resource}. The required number of shots then scales as
\begin{equation}
    N_{\mathrm{s}} = \tilde{\mathcal{O}}\left(p_q^{-2 \ln (|Q|^{-1})} \epsilon^{-2}\right).
\end{equation}
\end{lemma}

For a proof, please refer to Appendix~\ref{appendix:proofs}. Therefore, the total time complexity of our approach is obtained as the number of shots multiplied with the circuit depth from \cref{theo:joint-bound} -- we also compare this complexity with state-of-the-art statistical phase estimation techniques in Appendix~\ref{appendix:related}.

This implementation is well suited for modular quantum architectures~\cite{jnane2022multicore,araki2025space},
given that the deepest part of the algorithm is the time evolution, which can be performed independently within each module; then a relatively shallow and sparse derangement circuit is used to entangle the copies.
However, this implementation does require $n$ quantum registers to store the $n$ copies.

In the following, we detail another approach, in which only $1$ register is required, albeit the classical post-processing cost is substantially increased. Still, the reduction in the required number of qubits is worth the trade-off, especially in an early fault-tolerant setting where the number of logical qubits may be significantly limited. We highlight this trade-off by taking a surface code architecture as an example. In the implementation depicted in \cref{fig:vd_circuit}, independent time evolutions are performed on $n$ copies of the initial state before virtual distillation is applied. If we instead use only a single copy and repurpose the physical qubits to increase the size of the surface code logical qubits, then we can increase the code distance by a factor of $\mathcal{O}(\sqrt{n})$. This increase in code distance exponentially lowers the logical error rate and therefore enables a polynomially increased time evolution depth. The latter implies that $\sigma$ can be increased, which leads to an exponential reduction of algorithmic errors via \cref{lemma:time-evolution}.

\subsection{Suppressing non-dominant contributions by MC integration}
We now build on the observation that the non-linear functional $\tr[ \ravg^n O]$ 
of time-averaged states can be related to a high-dimensional integral.
This then leads to an approach that does not require $n$ copies of the average, mixed quantum state, but rather
estimates temporal correlators, which are then fed into a classical post-processing algorithm
to approximately evaluate a high-dimensional integral.

\begin{proposition}
\label{theo:main}
	The non-linear functional of time-averaged states $\ravg$ is equivalent to an $n$-dimensional integral over $\bm{t}=(t_1, t_2, \dots t_n)$ as
	\begin{align*}
		\tr[ \ravg^n O]
		&=\int    G(\bm{t}) F(\bm{t}) \,  \mathrm{d}\bm{t},
	\end{align*}
	where $G(\bm{t})$ is the PDF of an $n$-variate normal distribution
	and $F(\bm{t}) = A(t_n,t_1) \prod_{k=1}^{n-1}  B(t_k,t_{k+1})$
	depends only on the correlators $A(t,t') = \bra{\psi(t)} O \ket{\psi(t')}$ and $B(t,t') = \bra{\psi(t)} \ket{\psi(t')}$ for all $t$ and $t'$.
\end{proposition}
Please refer to Appendix~\ref{appendix:proofs} for a proof. One can apply quadrature integration formulas, which we also discuss in Appendix~\ref{appendix:quadrature}, to compute this integral numerically on a truncated discrete time grid. However, as the number of copies increases, this approach becomes highly inefficient as it requires resources that are exponential in the number of copies. Furthermore, certain quadrature techniques are highly susceptible to numerical instabilities in practical applications \cite{trefethen2022exactness}. Therefore, we focus on approximate integration via MC sampling that in turn provides unbiased estimates for the functionals in question.
\begin{lemma}
\label{statement:MCint}
	Given a quantum state $\ket{\psi(0)}$, the integral in \cref{theo:main} can be replaced by the random variable
	$F(\hat{ \bm{t}})$, where we choose $\bm{t}=(t_1, t_2, \dots t_n)$ according to
	the $n$-variate normal distribution with $G(\bm{t})$ as the PDF. We obtain an unbiased estimator as
	\begin{equation*}
		\mathds{E}[ F(\hat{ \bm{t}})  ] = \tr[ \ravg^n O],
	\end{equation*}
	through averaging the MC samples
	\begin{align}\label{eq:sample}
		F(\bm{t})
		= A(t_n,t_1) \prod_{k=1}^{n-1}  B(t_k,t_{k+1}).
	\end{align}
	These samples can be computed classically using the two objects $A(t,t') = \bra{\psi(t)} O \ket{\psi(t')}$ and $B(t,t') = \bra{\psi(t)} \ket{\psi(t')}$ that are classically stored in a 2D time grid as we detail below.
\end{lemma}

The denominator in \cref{theo:joint-bound} is obtained as a special case
\begin{equation}
    \tr[ \ravg^n] =\int    G(\bm{t}) J(\bm{t}) \,  \mathrm{d}\bm{t},
\end{equation}
where $J(\bm{t}) = B(t_n,t_1) \prod_{k=1}^{n-1}  B(t_k,t_{k+1})$. 
Evaluating these two integrals then allows us to estimate the expectation value of interest $\bra{\psi_q}O\ket{\psi_q}$ by estimating the ratio of two non-linear functionals $\tr[ \ravg^n O]$ and $\tr[ \ravg^n]$.

\subsection{Variance of the estimator}
\label{sec:variance}
We now use median of means estimators to derive rigorous performance guarantees. In the following, by each MC sample, we refer to one classical evaluation of the object $F(\hat{ \bm{t}})$ or $J(\hat{ \bm{t}})$. Using $2K$ batches of $N_{\mathrm{batch}}$ samples each, we obtain two estimators for the numerator $F$ (that estimates $\tr[ \ravg^n O]$ ) and denominator $J$ (that estimates $\tr[ \ravg^n]$ ) as

\begin{align}
    \hat{\mu}_{K,b}(F) \coloneqq \mathrm{median}\{ \hat{F}_1, \dots, \hat{F}_K\}, \\
    \hat{\mu}_{K,b}(J) \coloneqq \mathrm{median}\{ \hat{J}_1, \dots, \hat{J}_K\}. \nonumber
\end{align}

Using these two unbiased estimators, the error of their ratio can be bounded as described in the following theorem.

\begin{theorem}

    Two median of means estimators, which use $2KN_{\mathrm{batch}}$ independent samples in total, enable accurately predicting the ratio $\mathds{E}[F]/\mathds{E}[J]$ with failure probability $\delta$ and precision $\epsilon$. The precision is bounded as
    \begin{equation}
    	\epsilon \leq 2\frac{|\mathds{E}[F]|+\mathds{E}[J]}{\mathds{E}[J]^2}
    	{N^{-1/2}_{\mathrm{batch}}} + \mathcal{O}(
    	{N^{-1}_{\mathrm{batch}}}).
    	\label{eq:error_of_ratio}
    \end{equation}
     In leading terms in $N_{\mathrm{batch}}$, we obtain the approximate sample complexity as 
	\begin{equation}
    K = -8\log{\delta} \mathrm{\ \ and\ \ }N_{\mathrm{batch}} \lessapprox 4\frac{(|\mathds{E}[F]|+\mathds{E}[J])^2}{\mathds{E}[J]^4} \epsilon^{-2}.
\end{equation}
Given a vanishing expectation value $ \mathds{E}[F] = \tr[\ravg^n O] \rightarrow 0$, the approximate sample complexity scales as $N_{\mathrm{batch}} = \tilde{\mathcal{O}}\left(p_q^{-2\ln{(|Q|^{-1})}} \epsilon^{-2}\right)$.
\label{thm:med-of-means}
\end{theorem}

For a proof, please refer to Appendix~\ref{appendix:proofs}.

While the assumption $|\mathds{E}[F]| \rightarrow 0$ can be seen as a limiting factor, we argue that it can be approximately fulfilled using the following technique.
We can enforce $  \tr[ \ravg^n O] \approx 0$
by linearly combining the observable with the identity operator as $O \rightarrow O - c\mathds{1}$ for a suitably chosen scalar $c$, which is equivalent to applying the transformation $A(t, t') \rightarrow A(t, t') - c B(t, t')$ to our data set. In practice, we find that choosing $c$ to minimize $|A(t, t')- c B(t, t')|$ also nearly minimizes the MC variance~\cite{cai2025biased}.
We also note that this technique is not specific to median of means estimation and can be applied to other MC estimators as well, including those relying on the sample mean.

\subsection{Discrete time integral}
So far, we assumed that $t$ and $t'$ are sampled from a continuous normal distribution in our MC approach. However, as we now detail, the time variables can be discretized given our correlators are Fourier band-limited functions of time. 
Specifically, the two-time correlators $A(t,t') = \bra{\psi(t)} O \ket{\psi(t')}$ and $B(t,t') = \bra{\psi(t)} \ket{\psi(t')}$ are linear combinations of the frequency components $e^{-i(E_j-E_k)t}$, where $E_k$ are eigenvalues of the Hamiltonian.
	Therefore, the largest frequencies that can appear are upper bounded via the
	spectral radius of the Hamiltonian as $\lVert \mathcal{H} \rVert_{\circ }= \max_{j,k} |E_j-E_k|$.
According to the Nyquist–Shannon sampling theorem~\cite{press2007numerical},
sampling the correlators on a uniform 2D time grid is informationally complete, provided the timestep $dt$ satisfies the Nyquist condition relative to the highest angular frequency, i.e., $dt \propto \lVert \mathcal{H} \rVert_{\circ }^{-1}$.
Thus, as long as the continuous time correlators are sampled according to this condition, they 
can be reconstructed uniquely, e.g., one may use Fourier interpolation methods to reconstruct
the correlators to an arbitrarily fine time resolution without additional quantum evaluations,
given band-limited functions are fully determined by their Nyquist-rate samples.
Furthermore, based on our error bound in \cref{theo:joint-bound}, the maximum evolution time $T$ can be upper bounded by a constant multiple of $\Delta^{-1}$. 
This follows from the rapid decay of the Gaussian distribution, making the truncation error negligible by choosing $T$ to be a small constant multiple of the standard deviation $\sigma$ (e.g., $5 \sigma$).

\subsection{Estimating correlators using a quantum computer}
\label{subsection: obtaining}

We consider a uniform discretization of $[-T,T]$ into $N_T = 2T/dt + 1$ grid points. Each entry of the grid storing the $B$ object can be computed using a Hadamard test circuit shown in \cref{fig:algo},
where the initial state $\ket{\psi(0)}=U_{\text{init}}\ket{0}$ 
is prepared using $U_{\text{init}}$ and then time evolutions with durations $-t$ and $t'$ are applied conditioned on the state of the ancilla qubit.
Measuring the ancilla qubit in the $X$ ($Y$) basis gives an unbiased estimator for the real (imaginary) part of $B(t,t')$.
The circuit to obtain $A(t,t')$ is then identical to that for $B(t,t')$, except that the observable is additionally applied conditioned on the state of the ancilla qubit in between the two controlled time evolutions.

Estimating all entries of the $B$ correlator matrix requires $N_{\mathrm{s}} N_T^2$ executions. The $A$ correlator matrix similarly requires $N_{\mathrm{s}} N_T^2$ executions if $O$ is implemented via block encoding and $N_{\mathrm{s}} N_T^2 \lVert \bm{\nu} \rVert_1^2$ executions when the individual Pauli terms in the observable are sampled via importance sampling. Here, $\lVert \bm{\nu} \rVert_1$ is the $\ell_1$ norm of the Pauli coefficient vector $\bm{\nu}$ and the number of shots $N_{\mathrm{s}} \leq \epsilon_{\mathrm{s}}^{-2}$ is determined by the required final precision $\epsilon_{\mathrm{s}}$.

These circuits can be optimized as detailed in Appendix~\ref{appendix:improvements} (summarized in \cref{fig: QC}), so we refer to them as an unoptimized baseline construction. Specifically, only $N_T-1$ entries of the 2D time grid are unique in $B$, since its value depends on the time difference $t' - t$ rather than their unique combinations. The latter property also holds when $O$ commutes with the problem Hamiltonian in $A$, in which case as low as $N_T$ entries are unique. Thus, the number of circuit executions can be fundamentally reduced to scale linearly in $N_T$. Moreover, since $A$ corresponds to an expectation value rather than a two-time correlator when $t=t'$, the diagonal entries of $A$ do not require a Hadamard test, and can instead be evaluated using ancilla-free time evolutions with durations up to $2T$.

\subsection{Summary of the DDE algorithm}
\label{subsection: hybrid}
Here we summarize the workflow of DDE, which begins with quantum computation, proceeds through classical MC sampling, and finally estimates the desired eigenstate property. The input consists of a Hamiltonian $\mathcal{H}$, an initial state $\ket{\psi(0)}$ that has a dominant overlap with the eigenstate of interest $\ket{\psi_q}$, and a target observable $O$. The output is an
estimate of $\bra{\psi_q}O\ket{\psi_q}$ via the ratio between MC estimators $\mathds{E}[ F ]$ and $\mathds{E}[ J ]$.
\begin{enumerate}
    \item \textbf{Quantum computation:}
    Choose temporal grid with cutoff $T$ (consistent with the Gaussian window width $\sigma$)
    and timestep $dt$. The correlators $A(t,t')$ and $B(t,t')$  are estimated using a quantum device by evaluating them at discrete $t$ and $t'$ points. Details are discussed in \cref{subsection: obtaining} and Appendix~\ref{appendix:improvements}. 
    
    \item \textbf{Generate classical MC samples:}
    Choose hyperparameters for classical post-processing, namely the number of (virtual) copies $n$
    and the number of MC samples $N_{ \rm MC}$. Calculate a discrete probability
    distribution by evaluating the PDF of the normal distribution $\mathcal{N}(0,\sigma)$
    over the grid $[-T,T]$. Generate two sets of random time values $\hat{\bm{t}}_F = (\hat{t}_1, \hat{t}_1, \dots \hat{t}_n)$ and $\hat{\bm{t}}_J = (\hat{t}_1, \hat{t}_1, \dots \hat{t}_n)$ according to the discrete probability distribution $N_{ \rm MC}$ times. Each MC sample $F$ and $J$ is computed as a product of $A$ and $B$ values at indices determined by the $\bm{t}_F$ and $\bm{t}_J$ samples, respectively, as in
    \cref{eq:sample}. 
    
    \item \textbf{Estimate eigenstate property:}
    The generated MC samples are used to estimate both $\mathds{E}[ F ]$ and $\mathds{E}[ J ]$ in \cref{statement:MCint}. The desired estimate of the target eigenstate property is obtained by taking the ratio of empirical means $\mathds{E}[ F ]/\mathds{E}[ J ]$, or via the ratio of median of means as detailed above.
\end{enumerate}

Please note that the procedure outlined here is a baseline that can be improved in several ways, as we showcase in the following sections with concrete examples.

\section{Demonstrations}
\label{sec:demo}

In the following, we demonstrate the efficacy of DDE in numerical experiments
using different Hamiltonians relevant in condensed matter and high-energy physics applications.
We employ several combinations of initial state preparation, time evolution simulation, and classical post-processing techniques. 

\subsection{Exact simulation}

\label{sec:exactsim}

\begin{figure*}
	\centering
	\includegraphics[width=.8\textwidth]{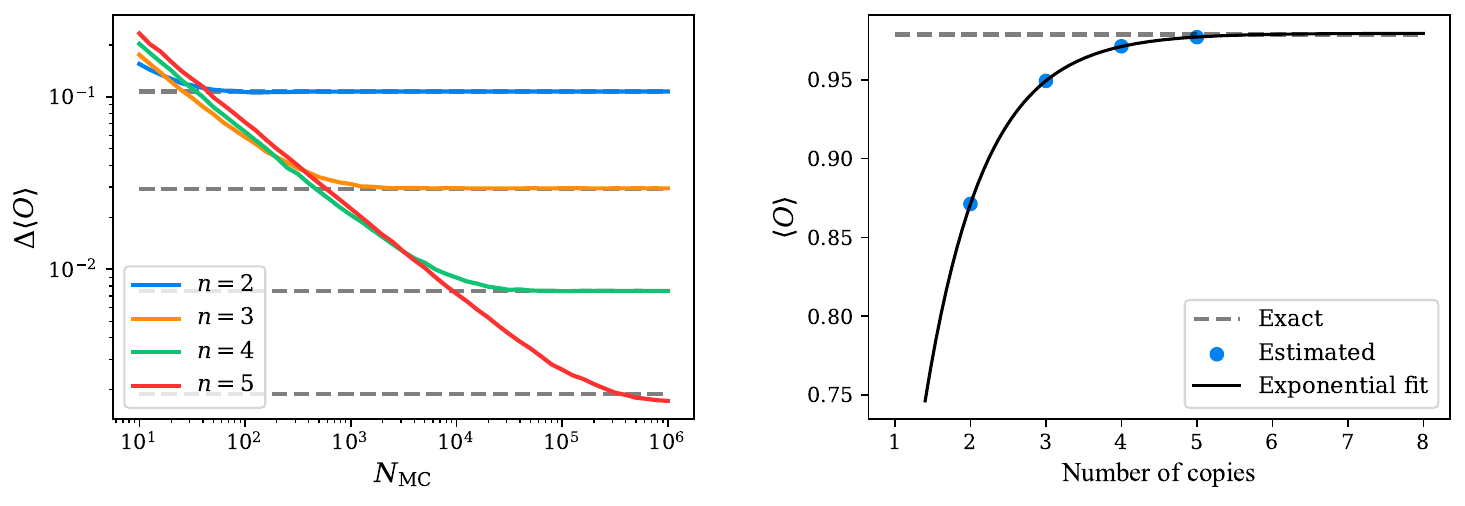}
	\caption{\textbf{Verifying DDE in a $10$-qubit random-field Heisenberg exact simulation.} (left) 
		Mean errors $\Delta\langle O \rangle$ (estimated from $10^4$ independent runs) for an increasing 
		number $n$ of virtual copies (solid lines) converge for an increasing number of MC samples to
		a systematic bias consistent with \cref{theo:joint-bound}. (right)
		Plotting the systematic bias for an increasing number $n$ of virtual
		copies and extrapolating to $n\rightarrow \infty$ via an exponential model function
		improves our estimate.}
	\label{fig: exact_spinring}
\end{figure*}

We first verify our DDE approach by performing exact time evolution in a random-field Heisenberg model on $10$ qubits with periodic boundary condition. Our initial state has $p_1=0.66$ overlap with the ground
state and $p_k=p_l=0.17$ with two other eigenstates, and we estimate the expectation value of $O=Z_1$.

We evaluate the correlators $A$ and $B$ over the time grid defined by $T = 200$ and $dt = 0.5$ numerically exactly, and use them in our MC approach.
In \cref{fig: exact_spinring} (left), we plot the mean error (solid lines) in our estimate of the expected value $\Delta\langle O \rangle$. The result confirms that, indeed, statistical fluctuations are suppressed, and the errors converge to the systematic bias (dashed lines) that we bounded in \cref{theo:joint-bound} as the number of MC samples is increased.
This systematic bias can be suppressed exponentially as we increase the number of virtual copies, and we achieve an error $10^{-2.76}$ with $5$ copies in the
present demonstration.

While the number of copies could be increased further, one can alternatively use converged values for up to 5 copies and apply extrapolation, as we demonstrate in \cref{fig: exact_spinring} (right). Specifically, we approximate our analytical model of the residual error in \cref{lemma:esd}
using an exponential function in $n$ given by $f(n) = a+cb^n$. We fit the parameters $a$ and $b$, and extrapolate to $n\rightarrow \infty$ to obtain an error $10^{-3.13}$ in the expectation value, which otherwise would require $6$ copies and $\approx 10^{7}$ MC samples. 

In practice, estimating the correlators $A$ and $B$ using a quantum device will introduce shot noise; however, in Appendix~\ref{appendix:numerics} we demonstrate that our MC approach is naturally robust against shot noise.

\subsection{Early fault-tolerant implementations}
Time evolution is one of the most natural applications of a quantum computer and is considered to be
the most promising area for achieving early quantum advantage. However, even the simplest 
systems that are challenging for conventional computers to simulate require quantum circuits potentially beyond the capabilities of non-error-corrected quantum devices~\cite{zimboras2025myths}. In the following, we illustrate how our approach can be implemented on early fault-tolerant quantum devices.

\subsubsection{Robustness to Trotter error}
\label{subsection: trotter error}

\begin{figure*}[t]
	\centering
	\includegraphics[width=.8\textwidth]{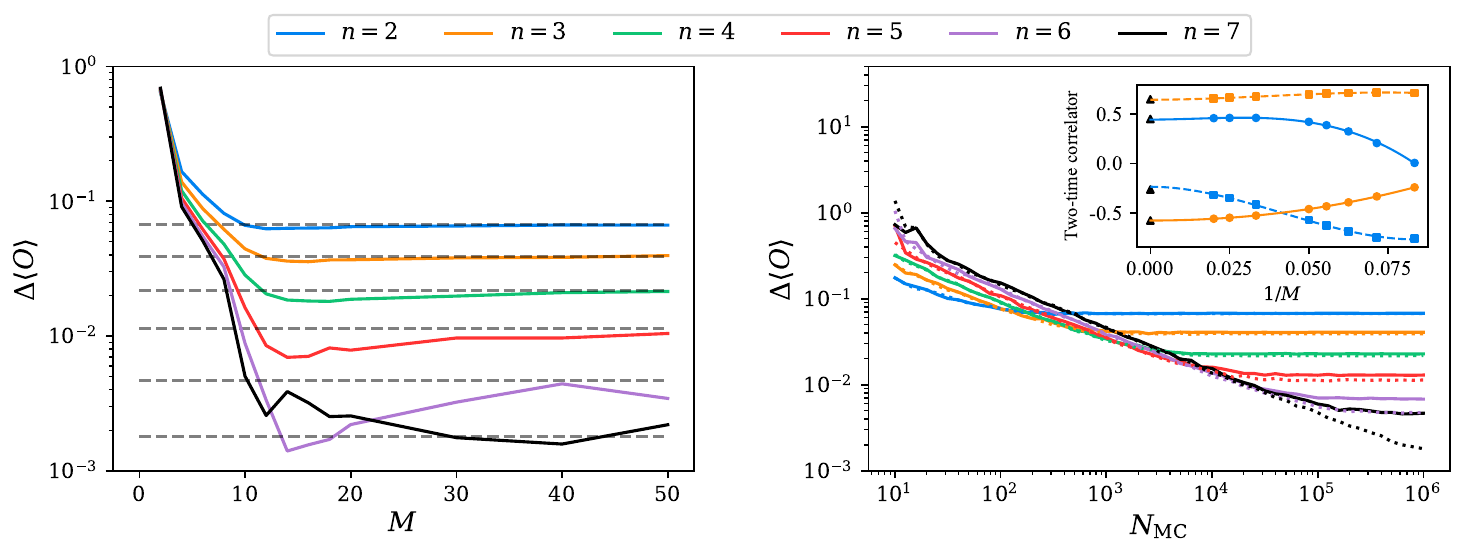}
	\caption{\textbf{Trotter simulation compared with exact time evolution in an $8$-qubit Fermi-Hubbard model.} (left)
		Mean errors in the estimated expectation value (using $10^6$ MC samples and averaging over $10^3$ runs)
		for an increasing number $M$ of Trotter steps and for an increasing number $n$ of copies.
		Mean errors using Trotterization (solid lines) rapidly (in $M$) converge
		to errors using exact time evolution (dashed lines) confirming the robustness of DDE to algorithmic errors.
		(right, inset) Applying polynomial extrapolation to the real (solid line, circles) and imaginary (dashed line, squares) parts
		of the single matrix entries $A(169.5, 93.5)$ (blue) and $B(-31.5, -70.5)$ (orange). The ideal values are also shown for comparison (black triangles).
		(right) Mean errors in the expected value estimation using correlators $A$ and $B$ estimated using 
		extrapolated entries (solid lines) closely approximate mean errors obtained via exact time evolution (dotted lines)
		confirming the efficacy of extrapolation as an algorithmic error mitigation technique.
	}
        
	\label{fig:trotter_steps}
\end{figure*}

It is expected that early fault-tolerant machines will enable simple Hamiltonian simulation algorithms
such as Trotterization~\cite{trotter1959product,suzuki1976generalized}.
While advanced variants can achieve zero algorithmic error~\cite{kiumi2024te},
product formulas in early or partially fault-tolerant devices will likely be limited by circuit depth and
therefore will have substantial algorithmic Trotter errors associated with the limited circuit depth. 
Here, we demonstrate the robustness of DDE to Trotter error and also demonstrate how such errors can be mitigated by applying straightforward extrapolation to the estimated $A$ and $B$ correlators.

We consider the Jordan-Wigner transformed Hamiltonian $\tilde{\mathcal{H}}$ of the Fermi-Hubbard model
such that a 4-site system is mapped to 8 qubits,
as described in Appendix~\ref{appendix:hubbard}, and drop the identity term in $\tilde{\mathcal{H}}$ to obtain $\mathcal{H} = \sum_j \nu_j P_j$ as a linear combination of $N_P$ Pauli terms $P_j$ with prefactors $\nu_j$.
We prepare the initial state $\frac{1}{\sqrt{2}}(\ket{10100101}+\ket{01011010})$, which has an overlap of $p_q \approx 0.62$ with the ground state (assuming half filling).
Time evolutions generated by $\mathcal{H}$ are applied on a time grid defined by
$T = 200$ and $dt = 0.5$. 

First-order Trotterization is implemented by further dividing each timestep into $M$ uniform intervals such that
\begin{equation}
	\label{eq:trotter}
	\exp{-i\mathcal{H}dt}
	\approx \bigg{(}\prod_j \exp{-i \nu_jP_j dt/M}\bigg{)}^M. \nonumber
\end{equation}
We consider a noiseless implementation of each Trotter step, such that the only source of error is the algorithmic Trotter error and choose the Pauli correlation $O=Z_1Z_2$ as our observable.
\cref{fig:trotter_steps} (left) shows the error in the estimated observable expectation value for an increasing number $M$ of Trotter steps and for an increasing number $n$ of virtual copies and confirms that the approach is indeed robust against algorithmic errors as the Trotterized time evolutions rapidly (in $M$) converge to exact ones (horizontal lines).

Given that the correlators $A$ and $B$ are estimated expectation values, algorithmic error mitigation techniques
can be applied~\cite{PRXQuantum.6.010352, rendon2024improved}. Here, we consider a
polynomial extrapolation of the Trotter error whereby we fit a degree-3 polynomial as $f(x)=ax^3+bx^2+c$
to all entries of $A$ and $B$ where $x$ is related to the Trotter step through $x = 1/M$.
We illustrate this extrapolation on single entries of $A$ (blue) and $B$ (orange) in \cref{fig:trotter_steps} (right, inset).
\cref{fig:trotter_steps} (right) verifies that DDE expectation values obtained via extrapolated $A$ and $B$ correlators (solid lines) closely approximate those obtained via exact time evolutions (dotted lines). This confirms that DDE is indeed compatible with error mitigation techniques.

\subsubsection{Robustness to gate noise}
\label{subsection: gate noise}

Early fault-tolerant devices will likely be limited by circuit depth due to non-negligible residual logical errors. Here, we demonstrate that DDE is robust against such errors. Our error model assumes that continuous rotation gates are the dominant source of errors, as these are fundamentally expensive components in fault-tolerant architectures~\cite{kliuchnikov2023shorter,PRXQuantum.5.040352,PRXQuantum.6.010352,goh2024direct}. For example, when rotations are synthesized using Clifford+T sequences, T gates are typically the most expensive components and more noise-prone than Clifford operations, as we detail in Appendix~\ref{appendix:early}.

In \cref{fig:noisy_trotter_steps}, we report simulations assuming two error severities such that the average number of errors in the time evolution circuit is $0.1$ (dotted lines) and $0.01$ (dashed lines) by assuming logical errors after each continuous rotation gate with probability $\gamma = 1/(100 N_G)$, and $\gamma = 1/(10 N_G)$, respectively, where $N_G = N_T N_P M$ is the total number of rotation gates in the Trotter circuits. Indeed, our DDE approach is robust to both error severities given the observable expectation values estimated with noisy circuits (dotted and dashed lines) closely approximate the estimation errors achieved with noise-free circuits (solid lines). Given that, in \cref{fig:noisy_trotter_steps}, we choose the observable to be the Hamiltonian as $O = \mathcal{H}$, we can exploit that $O$ commutes with the time evolution operator and therefore our quantum circuits can be significantly simplified (see Appendix~\ref{appendix:improvements}). Indeed, the optimized circuits yield smaller errors \cref{fig:noisy_trotter_steps} (right) than using the unoptimized circuits \cref{fig:noisy_trotter_steps} (left) for two reasons. First, reduced circuit depths guarantee less noise accumulation and second, shorter time evolution durations guarantee less Trotter algorithmic errors.

\begin{figure*}[t]
	\centering
	\includegraphics[width=.8\textwidth]{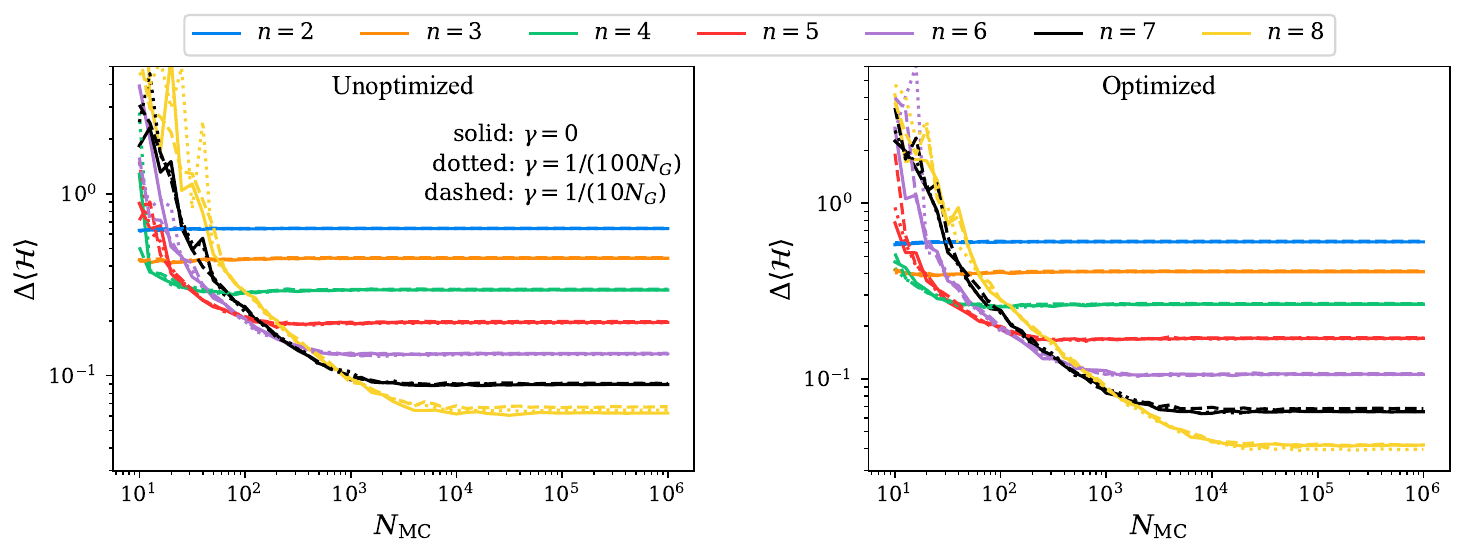}
	\caption{\textbf{Noisy Trotter circuit simulation of an $8$-qubit Fermi-Hubbard model}. 
	The mean error (over $10^3$ independent runs) of the observable ($O= \mathcal{H}$) expectation value	estimate for an increasing  number of MC samples is shown for increasing logical error rates (dashed vs dotted) 
		for both unoptimized (left) and optimized (right) circuits. The DDE approach is indeed robust against gate
		errors, given that the observable estimates using noisy circuits (dotted and dashed) closely approximate the noise-free
		circuits (solid lines).
		}     
	\label{fig:noisy_trotter_steps}
\end{figure*}

\subsection{Variational quantum simulation}

While fault-tolerant machines are emerging, it is broadly expected that the earliest forms of practical quantum advantage may be achieved with non-error-corrected devices in simulating the time evolution of relatively simple quantum systems, with variational simulation techniques~\cite{zimboras2025myths} considered to be the most promising approach. Here we consider a simple example from high-energy physics, the lattice Schwinger model \cite{hamer1997series}, as formulated in Ref.~\cite{chakraborty2022classicallyemulateddigitalquantum}. Our goal is to estimate the ground state energy of the corresponding Hamiltonian on $6$ qubits, described in Appendix~\ref{appendix:hep}.

\begin{figure*}[t]
	\centering
	\includegraphics[width=.8\textwidth]{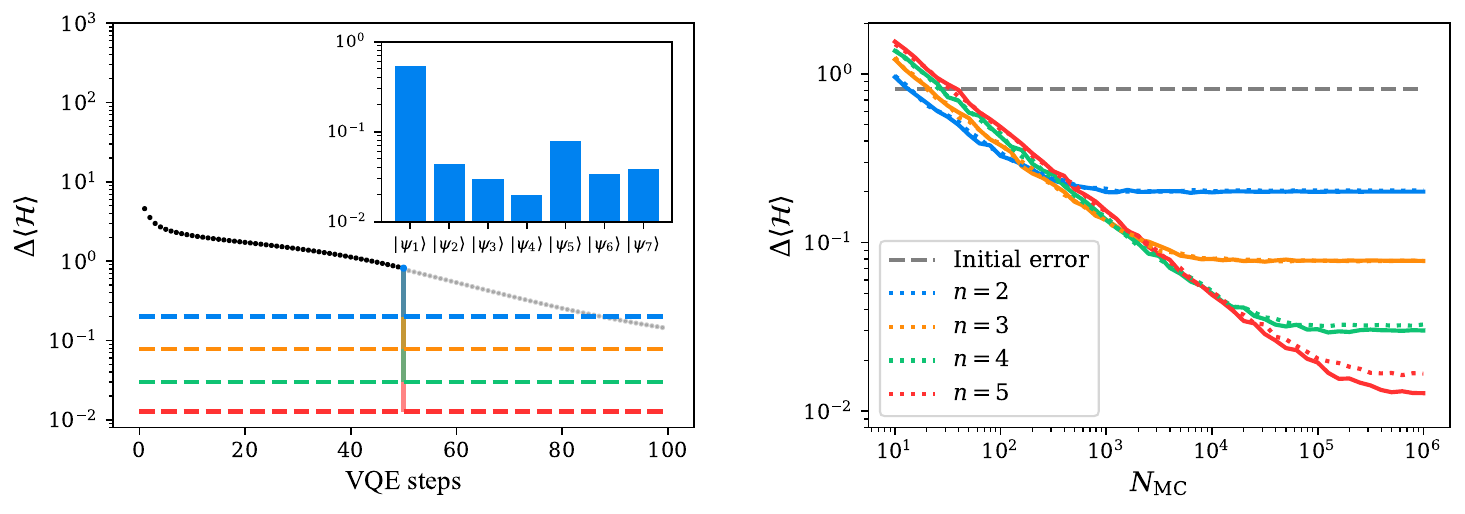}
	\caption{ \textbf{$6$-qubit lattice Schwinger model using variational time evolution.} (left) An initial state is obtained
		for DDE by stopping VQE (black dots) after only $50$ gradient descent steps. Applying DDE (dashed lines)
		improves exponentially on the estimation error of VQE as we increase the number of virtual copies. Inset: overlaps $|\braket{\psi_k}{\psi(\boldsymbol{\theta})}|^2$ between the DDE initial state  (obtained after 50 VQE iterations) and the lowest energy eigenstates $k \in [1,7]$. The initial state has an overlap of $p_1 = 0.53$ with the ground state and smaller overlaps with excited states. (right) Mean observable expectation value errors ($\Delta \langle\mathcal{H}\rangle$) for an increasing number of MC samples and number of copies. Dotted lines show results obtained from exact, while solid lines show the results from variationally simulated time evolution. The initial deviation from the exact ground state energy is also shown with the dashed horizontal line.}
	\label{fig:varsim_results}
\end{figure*}

First, we use VQE to obtain an initial state by optimizing parameters in an ansatz circuit $\ket{\psi(\boldsymbol{\theta})} = U(\boldsymbol{\theta})\ket{0}$ using gradient descent and deliberately terminate the training before convergence (after $50$ steps).
Here, we apply $12$ layers of a hardware-efficient ansatz with ring topology and show the training curve in \cref{fig:varsim_results} (left), as well as populations of the $7$ lowest energy eigenstates in the state obtained after $50$ training steps. Indeed, a significant disadvantage of VQE is that 
it is prone to getting stuck in local minima~\cite{anschuetz2022quantum}, which feature we exploit in the present application: 
The quantum state we obtain after only $50$ training steps has a sufficiently high $p_1=0.53$ overlap with the ground state $\ket{\psi_1}$
such that it can be used as an initial state for DDE.

We choose a main time grid defined by $T=50$ and $dt=0.2$ and apply ansatz-based variational time evolution~\cite{li2017efficient, yuan2019theory} using an integration timestep of $dt/1000$ to guarantee a sufficiently high fidelity -- the fidelity between the variationally time-evolved state and the exactly time-evolved state is above $99.4\%$ for all timesteps (as shown in \cref{fig:vqe_fidelity}). For further details on the ansatz-based variational quantum simulation algorithm, we refer the reader to Appendix~\ref{appendix:varsim}.

\cref{fig:varsim_results} (left, dashed lines) shows the observable expectation value errors when DDE is applied on these variationally time-evolved states and confirms that as we increase the number of virtual copies, the VQE energy error can be improved exponentially.
\cref{fig:varsim_results} (right) shows the mean observable expectation value errors for an increasing number of MC samples. Indeed, DDE is robust against algorithmic errors introduced by variational time evolution, given that the observable estimation errors (solid lines) closely approximate the errors obtained using exact time evolution (dotted lines).

DDE in combination with variational techniques seems to be a promising approach for near-term quantum advantage for the following reasons. While VQE is a heuristic approach that may require exponential resources to converge to the ground state, it almost always gets stuck in a local minimum~\cite{anschuetz2022quantum}, and we observe that these local traps may have sufficient overlap with low-lying eigenstates in numerical simulations. Applying DDE to these locally trapped VQE states is guaranteed to exponentially increase the quality of the observable estimate as long as efficient time evolution is possible.
While variational time evolution can be conveniently combined with VQE without increasing the ansatz circuit depth, these techniques typically require a very large number of shots and fine control of rotation angles~\cite{PhysRevLett.132.130602} to guarantee sufficiently high time evolution fidelity. We also note that while VQE is limited to ground states (or to low-lying eigenstates~\cite{higgott2019variational}),
our DDE approach can be applied to any eigenstate.

\section{Quantum-inspired classical simulation}
\label{sec:classical}
\subsection{Approach}
As detailed in \cref{subsection: obtaining}, our approach uses a quantum computer to estimate the correlators $A(t,t') = \bra{\psi(t)} O \ket{\psi(t')}$ and $B(t,t') = \bra{\psi(t)} \ket{\psi(t')}$. However, in a broad range of practically relevant scenarios, these correlators can instead be obtained using classical simulation techniques,
bypassing the need for a quantum computer. For example, when the initial state can be represented efficiently as an MPS, time evolution can be implemented using the well-established time-evolving block decimation (TEBD) algorithm~\cite{vidal2004efficient, verstraete2004matrix}, and therefore the desired correlators can be obtained through contractions.

Indeed, tensor networks cannot represent general quantum states efficiently. However, as long as our physical Hamiltonian of interest admits a time evolution of the initial state that can be captured by a family of tensor network states, then DDE can be fully simulated classically. This also means that DDE can improve and broaden the class of classical simulation techniques. For example, while preparing ground states is well-established via, e.g., the density matrix renormalization group (DMRG) technique~\cite{white1992density}, preparing excited states and highly excited states is non-trivial using tensor networks.

In contrast, if DDE is supplied with an initial state that has dominant overlap with one of the excited states, then time-evolving this state allows us to estimate observable expectation values in the desired excited state without explicitly preparing it. We demonstrate this application in the following section. Indeed, time evolution may lead to an increase in entanglement of the quantum system and MPS simulation will ultimately be limited to a certain maximal evolution time, which then limits the achievable Gaussian window $\sigma$. This leads to a finite approximation error in our bounds in \cref{theo:joint-bound} and further suppressing the simulation
error then necessitates a quantum computer -- which, however, can efficiently simulate the time evolution of any physical Hamiltonian.

\begin{figure*}[t]
	\centering
	\includegraphics[width=.8\textwidth]{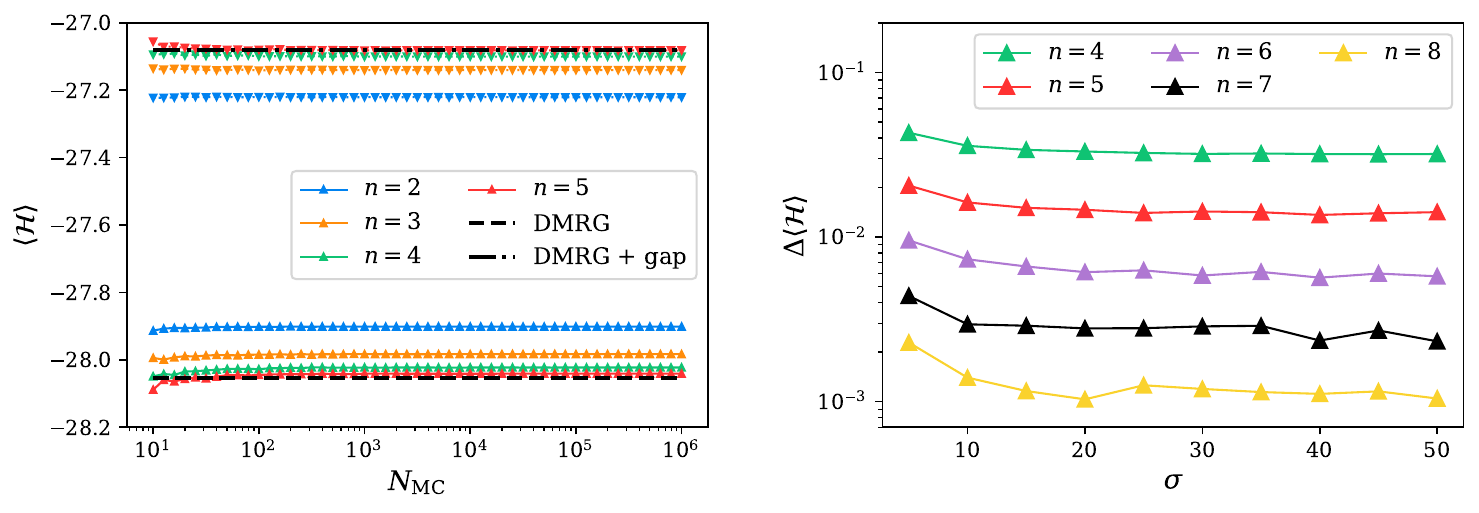}
	\caption{
		\textbf{MPS simulation of a $100$-qubit random-field Heisenberg model.} (left) Ground state and excited state energy estimates for an increasing number $n$ of copies and number of MC samples $N_{\mathrm{MC}}$ (triangles) converge to eigenenergy estimates obtained using alternative techniques --
		DMRG and spectroscopy (dashed lines). This confirms the efficacy of DDE which, however, is not limited to 
		energy estimation and can be applied to the 
		estimation of an arbitrary observable in the excited state -- whereas obtaining expected values in excited
		states is non-trivial using tensor-network techniques.		
		Here we also applied the variance reduction technique detailed in \cref{sec:variance}, i.e., the MC approach converges for
		relatively few samples. 
		(right) Our rigorous error bounds in \cref{theo:joint-bound} guarantee
		that the estimation error converges superexponentially as we increase the Gaussian time window $\sigma$,
		demonstrated here for an increasing number of copies using a fixed number $N_{\rm{MC}}=10^6$ (hence the non-smooth curves, and  $N_{\rm{MC}}=10^{6.5}$ for $n=8$). 
        Note that increasing $\sigma$ increases the depth of time evolution and thus may require an
		exponential increase in bond-dimension.
		\label{fig:mps_results}
	} 
\end{figure*}

\subsection{100-qubit MPS simulation \label{subsection: mps}}
We demonstrate our quantum-inspired classical DDE algorithm to obtain ground and excited state properties
of a $100$-qubit Heisenberg chain (with open boundary conditions) in an MPS simulation. 

We consider initial states of the form $\ket{\psi(0)} = c_1 \ket{b_1} + c_2 \ket{b_2}$, where the amplitudes $c_1$ and $c_2$ satisfy $\{|c_1|^2,|c_2|^2\} = \{0.7,0.3\}$ and $\{|c_1|^2,|c_2|^2\} = \{0.3,0.7\}$ for the ground and excited state demonstrations, respectively. $\ket{b_1}$ is the dominant bitstring in the DMRG ground state, and $\ket{b_2}$ is obtained by flipping
the single bit in $\ket{b_1}$ that has the least energy (approximation of a low-lying excited state).
We apply TEBD~\cite{vidal2004efficient, verstraete2004matrix} to these initial states
with a Trotter error bound $~10^{-4}$ at the maximal evolution time $T = 200$ using a step size $dt = 0.5$. 
We report the true ground and excited state energies in \cref{fig:mps_results} (left, dashed lines),
where the ground state energy was obtained using DMRG~\cite{white1992density}.  Our estimate of the
excited state energy was obtained through spectroscopy~\cite{PRXQuantum.6.010352}, whereby
we computed the overlaps $\braket{\psi(0)}{\psi(t)}$ for all discrete timesteps $t \in[0,200]$ whose Fourier transform is the local density of states \cite{ares2009loschmidt} and thus the frequency of the dominant peak corresponds to the energy of the excited state.

We then use the same time-evolved MPS states to estimate the correlators $A(t,t') = \bra{\psi(t)} O \ket{\psi(t')}$ and $B(t,t') = \bra{\psi(t)} \ket{\psi(t')}$, and apply DDE for estimating the expectation value of the Hamiltonian via $O=\mathcal{H}$. \cref{fig:mps_results} (left) shows the ground and excited state energies obtained using DDE for an increasing number of copies and for an increasing number of MC samples, and confirms that our approach indeed converges to the same energy estimates as DMRG and spectroscopy. Note that DDE is not limited to energy estimation and can be applied to estimate the expected value of any observable $O$ in the excited state that can be represented efficiently as a matrix product operator -- our motivation for estimating the energy is to verify DDE against established techniques.

Furthermore, \cref{fig:mps_results} (left) also nicely demonstrates our MC variance minimization approach 
detailed in \cref{sec:variance}, whereby we applied the transformation $A(t, t') \rightarrow A(t, t') - c B(t, t')$
via the coefficient $c = \bra{\psi(0)}\mathcal{H}\ket{\psi(0)}$.
This choice nearly minimizes the MC variance and, indeed, the MC estimates in \cref{fig:mps_results} (left) have sufficiently low uncertainty for a relatively low number of samples (please see more details in Appendix~\ref{appendix:mps_numerics}).

In \cref{fig:mps_results} (right), we also demonstrate the exponential suppression power for the ground state energy estimation task. This exponential suppression is consistent with our error bound in \cref{theo:joint-bound} and confirms the scalability of our approach to large system sizes well beyond exact diagonalization.

\section{Discussion and Conclusion}
\label{sec:discussion}

In this work, we introduced a hybrid quantum-classical algorithm, which enables the estimation of eigenstate properties by distilling the dominant contribution of an initial input state. Our DDE approach builds conceptually on purification-based QEM techniques applied to randomly time-evolved quantum states. In contrast to conventional VD, which uses multiple copies, our hybrid algorithm requires only a single quantum register with an ancillary qubit to estimate temporal correlators
over a 2D time grid. In the subsequent classical post-processing part, the exponential suppression power is achieved via a high-dimensional MC integration.

We presented rigorous theoretical guarantees for the algorithm, and proved that estimation errors are suppressed exponentially as we increase the number of (virtual) copies and super-exponentially as we increase the depth of time evolution (by increasing a Gaussian window width). Furthermore, we formally introduced performance guarantees associated with the MC-based classical post-processing step using median of means estimators.

We showcased the performance and scalability of our algorithm through an extensive set of numerical experiments in \cref{sec:demo}. We studied the robustness of DDE to both algorithmic errors associated with Trotterization and to residual logical errors in quantum gates as relevant in early fault tolerance. Furthermore, we demonstrated the compatibility of DDE with near-term-friendly variational techniques for both initial state preparation and time evolution.

Our approach is naturally compatible with a broad range of initial state preparation and time evolution techniques.
For example, while initial state preparation is in general exponentially expensive, for a broad range of practically important problems, good initialization is possible, e.g., by using tensor-network techniques~\cite{low2025fast, morchen2024classification}. We also demonstrated that VQE could be used to prepare an initial state for DDE. 
Furthermore, while time evolution is one of the most natural applications of quantum computers, current resource estimates suggest that even relatively simple systems may require deep time evolution circuits. Randomized time evolution formulas can significantly reduce absolute resource requirements for small to medium-scale systems and are indeed compatible with DDE~\cite{campbell2019random, kiumi2024te}. While DDE achieves shallow quantum circuits of depth $\tilde{\mathcal{O}}(\Delta^{-1})$, its runtime may become impractical when the spectral gap closes, as relevant, e.g., in glassy systems. This is, however, a general limitation in quantum computing that is consistent with the Quantum Merlin-Arthur (QMA) hardness of
eigenstate preparation \cite{bookatz2012qma, aharonov2003adiabatic}.
Still, most candidates for quantum advantage, such as FeMoco, target 
gapped Hamiltonians and are therefore well suited for our approach.
Furthermore, our gap dependence may be reduced significantly by
applying quadratic Hamiltonian factorization, which was originally introduced in
 the context of adiabatic computation, and recently applied to sum of squares simulation
 of the FeMoco Hamiltonian~\cite{somma2013spectral, low2017hamiltonian, zlokapa2024hamiltonian}.

Another promising direction may be to use analog quantum simulation as part of DDE, i.e., applying analog state preparation with low resource requirements (e.g., using adiabatic evolution) and then implementing analog time evolution. Correlators estimated from the analog quantum simulation experiments would then allow for a better estimate of the expectation value than with the simulation technique alone.

Finally, we also demonstrated that our approach is viable as a quantum-inspired classical simulation technique.
We demonstrated that DDE can be effectively combined with state-of-the-art classical tensor-network methods
to estimate both ground and excited state properties, thereby complementing the existing toolbox
of tensor-network simulation techniques.

\section{Acknowledgments}

The numerical simulation of quantum circuits made use of the Quantum Exact Simulation Toolkit (QuEST) \cite{jones2019quest} via the QuESTlink \cite{jones2020questlink} frontend, while tensor network simulations were performed with the quimb Python package \cite{gray2018quimb}. Quantum circuits were illustrated using the Quantikz LaTeX package \cite{kay2023tutorial}.
B.B. would like to thank the support of the Hungarian National Research, Development and Innovation Office (NKFIH) through the KDP-2023 funding scheme (grant number C2245275), and the TKP2021-NVA funding scheme (grant number TKP2021-NVA-29). B.B. also acknowledges the computational resources provided by the Wigner Scientific Computational Laboratory (WSCLAB). T.A. acknowledges support by the Oxford-Uehiro Graduate Scholarship Programme. TA also acknowledges the use of the University of Oxford Advanced Research Computing (ARC)
facility \cite{richards2015university} in carrying out this work.
B.K. thanks UKRI for the Future Leaders Fellowship Theory to Enable Practical Quantum Advantage (MR/Y015843/1).
The authors also acknowledge funding from the
EPSRC projects Robust and Reliable Quantum Computing (RoaRQ, EP/W032635/1)
and Software Enabling Early Quantum Advantage (SEEQA, EP/Y004655/1). B.K. thanks the University of Oxford for
a Glasstone Research Fellowship and Lady Margaret Hall, Oxford for a Research Fellowship.
This research was funded in part by UKRI (MR/Y015843/1).
For the purpose of Open Access, the author has applied a CC BY public copyright licence
to any Author Accepted Manuscript version arising from this submission.

\appendix

\section{Related works}
\label{appendix:related}

Algorithmic cooling techniques \cite{huo2023error,zeng2022universal} rely on quantum 
computation of the same two-time correlations as in our work, and classical computation of similar integrals through MC sampling to estimate eigenstate properties. While these prior works rely on imaginary time evolution, DDE is based on real-time evolution
and provably converges super-exponentially to a mixed state to which we apply virtual distillation -- which latter step provides a provably exponential error suppression power with respect to the number of virtual copies. These methods also rely on prior knowledge of the target eigenstate energy to alleviate the sampling overhead, similarly to the technique that we discuss in \cref{sec:variance}. While we choose to sample random time evolutions from a Gaussian distribution, these works have considered alternative distributions, which may also prove useful in the context of DDE.

Our work is also closely related to quantum virtual cooling \cite{cotler2019quantum}, which inputs a thermal state and obtains ground state properties via virtual purification. This is similar to applying DDE with the fully quantum implementation of controlled derangement, rather than offloading the computational burden to MC-based classical post-processing, as discussed in \cref{subsection: suppressing}. In our work, the thermal state is replaced by an average of randomly time-evolved states, which leads to the following advantages. First, we reduce the requirement of the number of quantum registers from $n$ to $1$. Second, we enable targeting higher excited states by appropriately choosing the initial state. Third, we provide an additional flexibility in how to perform initial state preparation and classical virtual purification.

Our approach shares similarities with statistical phase estimation techniques, i.e., it similarly requires
only a single ancilla qubit and relies on estimating similar objects for subsequent classical post-processing.
While a range of statistical phase estimation techniques are known that make a trade-off between maximal circuit depth and low number of shots,
remarkably, the circuit depth (maximum evolution time) of \dde \, matches the optimal $\tilde{\mathcal{O}}(\Delta^{-1})$ scaling of the algorithms introduced in Refs.~\cite{wang2023quantum, wang2025efficient} for energy estimation.

On the other hand, Refs.~\cite{wang2023quantum, wang2025efficient} report a scaling
of $\tilde{\mathcal{O}}(p_q^{-2}\epsilon^{-2})$ with respect to the number of shots but
require prior knowledge of a lower bound on the spectral gap $\Delta$ and the initial overlap $p_q$. 
While \dde \, can operate under more relaxed assumptions, it requires an increased number of shots $\tilde{\mathcal{O}}(p_q^{-2\ln{(|Q|^{-1})}}\epsilon^{-2})$. We argue below that a) the increased number of shots is inevitable, given \dde\ is fundamentally more general, and  b) that we consider practically important early fault-tolerant settings where $p_q \approx 1$, which then results in a reasonable number of shots as demonstrated in our numerical experiments.
As a sidenote, we also remark that the approach of Ref.~\cite{lin2022heisenberg} reaches Heisenberg scaling in the total runtime $\tilde{\mathcal{O}}(p_q^{-2}\epsilon^{-1})$, but it requires deeper circuits given its $\tilde{\mathcal{O}}(\epsilon^{-1})$ scaling.

Let us now illustrate why estimating an arbitrary expectation value in an eigenstate is a more complex task than estimating eigenenergies given access to an initial state~\cite{bookatz2012qma}.
Estimating eigenenergies with textbook phase estimation has a complexity of $\mathcal{O}(p_q^{-2} \epsilon^{-1})$~\cite{nielsen2010quantum, lin2022heisenberg, ge2019faster}.
In contrast, using phase estimation to prepare an eigenstate and then sampling to estimate an arbitrary expected value would require a significantly increased runtime $\mathcal{O}(p_q^{-2} \epsilon^{-3})$.
Remarkably, our sampling-based \dde \, matches the complexity of phase estimation combined with amplitude estimation~\cite{brassard2000quantum}
in its scaling with respect to the target precision, as the latter has a complexity of $\mathcal{O}(p_q^{-2} \epsilon^{-2})$.
We have an increased complexity with respect to the overlap in the initial state $\tilde{\mathcal{O}}(p_q^{-2\ln{(|Q|^{-1})}})$. However, as we detailed in \cref{sec:intro}, the most promising candidates for achieving early quantum advantages are problems where reasonably good initial states are known with $p_q \approx 1$, thanks to decades of classical advancements. In these cases, quantum computing is necessary to improve the precision to a practically useful level, as exemplified by FeMoco ($p_q \approx 0.9$).

\section{Proofs}
\label{appendix:proofs}
\subsection{Proof of \cref{lemma:time-evolution}}

\begin{proof}
Explicitly, the time-evolved state is
	\begin{equation}
		\rho(t) = \ketbra{\psi(t)} = \sum_{j,k} c_j c^*_k e^{-i t (E_j-E_k)} \ketbra{\psi_j}{\psi_k}.
	\end{equation}
	We can substitute this back into \cref{eq:t_integral},
	\begin{align}
		 \ravg
		&= \int_{-\infty}^{\infty} G(t)  \rho(t) \, \mathrm{d}t\\
		& = \sum_{j,k} c_j c^*_k \ketbra{\psi_j}{\psi_k}  \int_{-\infty}^{\infty} G(t)   e^{-i t (E_j-E_k)}  \, \mathrm{d}t. \nonumber
	\end{align}
	The integral above is the Fourier transform of the Gaussian distribution as
	\begin{equation}
		\int_{-\infty}^{\infty} G(t)   e^{-i t (E_j-E_k)}  \, \mathrm{d}t
		=
		e^{-\frac{(E_j-E_k)^2 \sigma^2}{2}}.
	\end{equation}
	Substituting this back yields
	\begin{align}
		\ravg
		& = \sum_{j,k} c_j c^*_k  e^{-\frac{(E_j-E_k)^2 \sigma^2}{2}} \ketbra{\psi_j}{\psi_k}  \\
		& = \underbrace{\sum_{k} |c_k|^2  \ketbra{\psi_k}{\psi_k}}_{\rho} 
		+  \underbrace{\sum_{j \neq k} c_j c^*_k  e^{-\frac{(E_j-E_k)^2 \sigma^2}{2}} \ketbra{\psi_j}{\psi_k}}_{\mathcal{E}}. \nonumber
	\end{align}
	Let us now bound the Hilbert-Schmidt distance $\lVert \mathcal{E} \rVert_{\mathrm{HS}} = \lVert \ravg - \rho \rVert_{\mathrm{HS}}$, exploiting the fact that the squared coefficients $p_k = |c_k|^2$ form a probability distribution,
    \begin{align}
		\lVert \mathcal{E} \rVert_{\mathrm{HS}}^2
		&= \left\lVert \sum_{j \neq k} c_j c^*_k  e^{-\frac{(E_j-E_k)^2 \sigma^2}{2}} \ketbra{\psi_j}{\psi_k} \right\rVert_{\mathrm{HS}}^2\\
		&= \sum_{j \neq k} |c_j|^2 |c_k|^2  \big (e^{-\frac{(E_j-E_k)^2 \sigma^2}{2}} \big )^2 \nonumber\\
		&\leq e^{-\Delta^2 \sigma^2} \sum_{j \neq k} |c_j|^2 |c_k|^2 \nonumber \\
		&= e^{-\Delta^2 \sigma^2} \left(\sum_{j,k} p_j p_k - \sum_k p_k^2 \right) \nonumber\\
		&=e^{-\Delta^2 \sigma^2} \left( 1 - e^{-H_2(\bm{p})} \right) \nonumber\\
            & \leq e^{-\Delta^2 \sigma^2} \left( 1 - (\max\limits_{k} p_k)^{2} \right). \nonumber
    \end{align}
Above we introduced the smallest gap in the spectrum of the Hamiltonian with non-zero contribution in the initial state, i.e., $\Delta = \min\{ |E_j-E_k| \; \mathrm{with} \; j,k:  |c_j|,|c_k|> 0  \}$. 

Next, we bound $\|\mathcal{E}\|_1$, the trace distance between the average state and the diagonal state. This can be written as the sum of vector norms 
\begin{equation}
    \| \mathcal{E} \|_1 = \sum_l \| \mathcal{E} \ket{x^l} \|_{2}.
    \label{eq:trace_to_vector_norm}
\end{equation}
Here $\ket{x^l}$ is the $l$th eigenvector of $\mathcal{E}$ with eigenvalue $\lambda_l$, therefore, due to the eigenvalue equation, we have $\| \mathcal{E} \ket{x^l} \|_{2} = |\lambda_l|$. 

With the notation $g_{jk} = \exp(-(E_j-E_k)^2 \sigma^2/2)$, each term of the sum in \cref{eq:trace_to_vector_norm} can be written as 
\begin{align}
    \| \mathcal{E} \ket{x^l} \|_{2}^2 & = \sum_j \left| \sum_{k\neq j} c_j c_k^* \,\, g_{jk} \, x_k^l \right|^2 \\
    & = \sum_j |c_j|^2 \left( \sum_{k\neq j} c_k^* g_{jk} x_k^l \right) \left( \sum_{k'\neq j} c_{k'} g_{jk'}^* (x_k^l)^* \right) \nonumber \\
    & = \sum_j |c_j|^2 | \bra{x^l}W_j\ket{\bm c}|^2, \nonumber
\end{align}
where $\bm c = (c_1, c_2 , \dots)$, and we introduced the diagonal matrices $W_j = \mathrm{diag}(g_{j1}, g_{j2}, \dots)$. With this, the trace norm of the error term can be bounded as
\begin{align}
    \|\mathcal{E}\|_1 & = \sum_l \sqrt{\sum_j |c_j|^2 | \bra{x^l}W_j\ket{\bm c}|^2} \\
    & \leq e^{-\Delta^2\sigma^2/2} \sum_l |\bra{x^l}\ket{\bm c}|. \nonumber
\end{align}
Similarly to the case of the Hilbert-Schmidt norm, here we also used the fact that $(W_j)_{kk} = g_{jk} \leq \exp(-\Delta^2 \sigma^2/2)$. The factor $\sum_l |\bra{x^l}\ket{\bm c}| = \| \bm c\|_1$ is the $\ell_1$ norm of the coefficient vector $ \bm c$ in the eigenbasis of the residual operator $\mathcal{E}$. The vector norm $\lVert \bm c \rVert_2=1$ is bounded and basis-independent, but the $\ell_1$ norm is basis-dependent and therefore
can in principle be as large as $\sqrt{2^N}$. 
However, our numerical demonstrations point to the fact that it remains within a reasonable scale in practice. 
This shows that the trace norm of the error term also vanishes super-exponentially with the standard deviation of the normal distribution. 
\end{proof}

\subsection{Proof of \cref{theo:joint-bound}}
\begin{proof}
    The quantity of interest is
    \begin{equation}
        |\mathcal{Q}| = \left|\frac{\mathrm{tr}[\ravg^nO]}{\mathrm{tr}[\ravg^n]} - \bra{\psi_q} O \ket{\psi_q}\right|,
    \end{equation}
    which can be bounded as
    \begin{equation}\label{eq: t1.1}
        |\mathcal{Q}| \leq \left|\frac{\mathrm{tr}[\ravg^nO]}{\mathrm{tr}[\ravg^n]} - \frac{\mathrm{tr}[\rho^nO]}{\mathrm{tr}[\rho^n]}\right| + \left|\frac{\mathrm{tr}[\rho^nO]}{\mathrm{tr}[\rho^n]} - \bra{\psi_q} O \ket{\psi_q}\right|
    \end{equation}
    by the triangle inequality. The second term is simply the error magnitude $|\mathcal{A}|$ from \cref{lemma:esd}. To bound the first term, we use the inequality from Ref.~\cite{koczor2021dominant} as
    \begin{equation}\label{eq: t1.2}
        \left|\frac{\mathrm{tr}[\ravg^nO]}{\mathrm{tr}[\ravg^n]} - \frac{\mathrm{tr}[\rho^nO]}{\mathrm{tr}[\rho^n]}\right| \leq 2 \|O\|_{\infty} \left\lVert \frac{\ravg^n}{\mathrm{tr}[\ravg^n]} - \frac{\rho^n}{\mathrm{tr}[\rho^n]} \right\rVert_1.
    \end{equation}
    Denoting deviations $\Omega$ and $\omega$ such that $\ravg^n = \rho^n + \Omega$ and $\tr[\ravg^n] = \tr[\rho^n] + \omega$, a further use of the triangle inequality leads to
    \begin{align}
\label{eq: t1.3}
        \bigg\lVert \frac{\ravg^n}{\mathrm{tr}[\ravg^n]} -& \frac{\rho^n}{\mathrm{tr}[\rho^n]} \bigg\rVert_1 = \left\lVert \frac{\rho^n + \Omega}{\mathrm{tr}[\rho^n] + \omega} - \frac{\rho^n}{\mathrm{tr}[\rho^n]} \right\rVert_1 \\
        \leq& \left\lVert\frac{\rho^n}{\mathrm{tr}[\rho^n]+\omega} - \frac{\rho^n}{\mathrm{tr}[\rho^n]} \right\rVert_1 + \left\lVert \frac{\Omega}{\mathrm{tr}[\rho^n]+\omega}\right\rVert_1. \nonumber
    \end{align}
    Both of these terms can be treated using the Maclaurin series expansion
\begin{equation}
\label{eq: t1.4}
    \frac{1}{\mathrm{tr}[\rho^n]+\omega} = \sum_{k = 0}^\infty (-1)^k \frac{\omega^k}{\mathrm{tr}[\rho^n]^{k+1}}.
\end{equation}
For the first term in \cref{eq: t1.3}, applying the triangle inequality to the expansion gives
\begin{align}\label{eq: t1.5}
    \left\lVert\frac{\rho^n}{\mathrm{tr}[\rho^n]+\omega} - \frac{\rho^n}{\mathrm{tr}[\rho^n]} \right\rVert_1 & \leq \left\lVert \frac{\rho^n}{\mathrm{tr}[\rho^n]} \right\rVert_1\sum_{k = 1}^\infty \left\lvert\frac{\omega}{\mathrm{tr}[\rho^n]}\right\rvert^k \nonumber \\
    & \leq \sum_{k = 1}^\infty \bigg{(}\frac{\left\lVert\Omega\right\rVert_1}{p_q^n}\bigg{)}^k,
\end{align}
where we also used the fact that $\omega = \mathrm{tr}[\Omega]$ implies $|\omega| \leq \lVert \Omega \rVert_1$. The second term in \cref{eq: t1.3} can be bounded similarly as
\begin{equation}\label{eq: t1.6}
        \left\lVert \frac{\Omega}{\mathrm{tr}[\rho^n]+\omega}\right\rVert_1 \leq \sum_{k = 0}^\infty \left\lVert \frac{\omega^k\Omega}{\mathrm{tr}[\rho^n]^{k+1}}\right\rVert_1 \leq \sum_{k = 1}^\infty \bigg(\frac{\lVert \Omega \rVert_1}{p_q^n}\bigg)^k.
    \end{equation}
Thus, both terms are bounded by the same infinite series with $\lVert \Omega \rVert_1/p_q^n$ as the common ratio. We proceed by expanding $\bar{\rho}^n = (\rho + \mathcal{E})^n$, which leads to
\begin{align}
\label{eq: t1.7}
        \lVert \Omega \rVert_1 & = \bigg\lVert \sum_{k=0}^{n-1} \rho^k\mathcal{E}\rho^{n-1-k} + \sum_{\substack{k,l = 0 \\ l+k\leq n-2}}^{n-2}\rho^k\mathcal{E} \rho^l \mathcal{E}\rho^{n-2-k-l} \nonumber \\
        & + \ldots + \sum_{k=0}^{n-1} \mathcal{E}^k \rho \mathcal{E}^{n-1-k} + \mathcal{E}^n \bigg\rVert_1\\
        & \leq \sum_{k=0}^{n-1} \left\lVert \rho^k\mathcal{E}\rho^{n-1-k} \right\rVert_1+ \sum_{\substack{k,l = 0 \\ l+k\leq n-2}}^{n-2} \left\lVert \rho^k\mathcal{E} \rho^l \mathcal{E}\rho^{n-2-k-l} \right\rVert_1 \nonumber \\ 
        & + \ldots + \sum_{k=0}^{n-1}\left\lVert \mathcal{E}^k \rho \mathcal{E}^{n-1-k} \right\rVert_1+ \left\lVert \mathcal{E}^n \right\rVert_1 \nonumber \\
        & \leq \sum_{k=1}^{n} {n \choose k} p_q^{n-k}\lVert \mathcal{E} \rVert_1^k. \nonumber
    \end{align}
The first inequality follows from the triangle inequality. The second inequality applies H\"{o}lder's inequality, the monotonicity of Schatten norms, and the fact that $\lVert \rho^k\rVert_{\infty} = p_q^k$ for all $k$. This gives an upper bound on the base of the infinite series in \cref{eq: t1.5} and \cref{eq: t1.6} as
\begin{equation}\label{eq: t1.8}
    \frac{\lVert \Omega \rVert_1}{p_q^n} \leq \sum_{k=1}^{n} {n \choose k} p_q^{-k}\lVert \mathcal{E} \rVert_1^k = (1 + \lVert\mathcal{E}\rVert_1/p_q)^n - 1,
\end{equation}
so that \cref{eq: t1.3} implies
\begin{align}
    \left\lVert \frac{\ravg^n}{\mathrm{tr}[\ravg^n]} - \frac{\rho^n}{\mathrm{tr}[\rho^n]} \right\rVert_1 & \leq 2 \sum_{k = 1}^\infty \bigg(\frac{\lVert \Omega \rVert_1}{p_q^n}\bigg)^k \\
    & \leq 2np_q^{-1}\lVert \mathcal{E} \rVert_1 + \mathcal{O}(\lVert \mathcal{E} \rVert_1^2). \nonumber
\end{align}
Piecing all the bounds together through \cref{eq: t1.1} and \cref{eq: t1.2}, we finally obtain
\begin{align}
    |\mathcal{Q}| \leq |\mathcal{A}| + 4 n p_q^{-1} \| O \|_{\infty} \| \mathcal{E} \|_1 + \mathcal{O}(\|\mathcal{E}\|_1^2).
\label{eq:app-bound}
\end{align}
Note that the above expressions assume that we are interested in the regime of vanishing $\lVert \mathcal{E} \rVert_1$, but higher-order terms in the upper bound can also be written out exactly based on \cref{eq: t1.8}. While these higher-order terms involve binomial coefficients, the bound
\begin{equation}
    \sum_{k=1}^{n} {n \choose k} p_q^{-k}\lVert \mathcal{E} \rVert_1^k < \sum_{k=1}^{n} \bigg{(}\frac{en}{p_q}\lVert \mathcal{E} \rVert_1\bigg{)}^k
\end{equation}
makes it clear that $\lVert \mathcal{E} \rVert_1 < p_q /(en)$ serves as a sufficient condition for the first-order term to be the leading term, where $e$ is the Euler number.

Finally, let us use the bound from \cref{eq:app-bound} to asymptotically estimate the resource requirements. Without loss of generality, we assume
\begin{align}
    |\mathcal{Q}|/2 & \leq 2 (p_q^{-1} - 1)^n e^{-(n-1)H_n(\bm p')}.
\end{align}
After using the monotonicity of R\'enyi entropies and rearranging the inequality, we get
\begin{equation}
\label{eq:n_bound}
    n \leq \frac{\ln[4|\mathcal{Q}|^{-1}]+H_{\infty}(\bm p')}{\ln[(p_q^{-1}-1)^{-1}]+H_{\infty}(\bm p')}.
\end{equation}
From this, we get the scaling $n = \mathcal{O}(\ln[|\mathcal{Q}|^{-1}])$.

Similarly, for the second term,
\begin{equation}
    |\mathcal{Q}|/2\lessapprox 4np_q^{-1} e^{\frac{-\Delta^2\sigma^2}{2}} \lVert\bm c\rVert_1 \lVert O \rVert_{\infty}.
\end{equation}
By rearranging this, we get
\begin{equation}
    \sigma \lessapprox \frac{\sqrt{2\ln[|\mathcal{Q}|^{-1}n] + 2 \ln[8 \lVert\bm c\rVert_1 \lVert O \rVert_{\infty}]}}{\Delta}.
\end{equation}
After Taylor expansion, we get the final scaling $\sigma = \mathcal{O}(\Delta^{-1}\sqrt{\ln[|\mathcal{Q}|^{-1}n]})$.
\end{proof}

\subsection{Proof of sampling overhead}
Consider applying DDE via direct estimation of non-linear functionals as discussed in \cref{subsection: suppressing}. The associated sampling overhead scales as $\mathcal{O}(\mathrm{tr}[\ravg^n]^{-2})$ \cite{koczor2021exponential,huggins2021virtual}, which is upper bounded as
\begin{align}
\begin{split}
    \mathrm{tr}[\ravg^n]^{-2} &= \mathrm{tr}[\rho^n]^{-2}\sum_{k = 0}^{\infty}(-1)^k(k+1)\bigg{(}\frac{\omega}{\mathrm{tr}[\rho^n]}\bigg{)}^k\\
    &\leq \mathrm{tr}[\rho^n]^{-2}\sum_{k = 0}^{\infty}(k+1)\bigg{(}\frac{\lVert\Omega\rVert_1}{\mathrm{tr}[\rho^n]}\bigg{)}^k\\
    &\leq p_q^{-2n}\big{(}1+2np_q^{-1}\lVert\mathcal{E}\rVert_1 + \mathcal{O}(\lVert\mathcal{E}\rVert_1^2)\big{)}.
\end{split}
\end{align}
We used the Maclaurin series expansion for the first equality. The triangle inequality and the fact that $|\omega|\leq \lVert\Omega\rVert_1$ are applied in the second step. The fact that $\mathrm{tr}[\rho^n]=\sum_k p_k^n \geq p_q^n$ and \cref{eq: t1.8} are applied in the third step.

Using the bound on the number of copies from \cref{eq:n_bound}, in leading terms, the sampling overhead can be written as $\tilde{\mathcal{O}}(p_q^{-2\ln{(|Q|^{-1})}} \epsilon^{-2})$, where $\epsilon$ is the standard statistical uncertainty. An important observation here is that the exponent depends only logarithmically on $|Q|^{-1}$, meaning that the complexity will be effectively polynomial in the inverse systematic bias.

\subsection{Proof of \cref{theo:main}}
\begin{proof}
	\begin{align}
		\ravg^n
		&= \Big( \int_{-\infty}^{\infty} G(t)  \rho(t) \, \mathrm{d}t \Big)^n\\
		&=  \int  G(t_1)  \rho(t_1) G(t_2)  \rho(t_2) \cdots G(t_n)  \rho(t_n)  \, \mathrm{d} \, t_1 \cdots \mathrm{d}t_n \nonumber\\
		&=\int  \Big( \prod_{k=1}^n G(t_k) \Big) \Big( \prod_{k=1}^n \rho(t_k) \Big) \,  \mathrm{d}t_1 \cdots \mathrm{d}t_n, \nonumber
	\end{align}
	thus 
	\begin{equation}
		\tr[ \ravg^n O]
		=\int  \Big( \prod_{k=1}^n G(t_k) \Big) \tr[ \rho(t_1) \cdots \rho(t_n) O] \,  \mathrm{d}t_1 \cdots \mathrm{d}t_n .
        \end{equation}
	Observing that the product in the expression above is a  multivariate Gaussian distribution $ \prod_{k=1}^n G(t_k) = G(\bm{t})$
	and introducing the notation for the trace as 
	$F(\bm{t}) = \tr[ \rho(t_1) \rho(t_2) \dots \rho(t_n) O] $
	we can compactly write the integral as 
	\begin{equation}
		\tr[ \ravg^n O]
		=\int    G(\bm{t}) F(\bm{t}) \,  \mathrm{d}\bm{t}.
	\end{equation}

	We can significantly simplify $F(\bm{t})$ by exploiting that $\rho(t_k) = \ketbra{\psi(t_k)}{\psi(t_k)}$ is a projection operator and using the cyclic property of the trace as
	\begin{align}
		F(\bm{t})
		=& 
		\bra{\psi(t_n)} O \ket{\psi(t_1)} \prod_{k=1}^{n-1} \bra{\psi(t_k)} \ket{\psi(t_{k+1})} \\
		=& A(t_n,t_1) \prod_{k=1}^{n-1}  B(t_k,t_{k+1}). \nonumber
	\end{align}
	
\end{proof}

\subsection{Proof of \cref{thm:med-of-means}}

\begin{proof}

The variance of the function $F(\bm{t})$ can be bounded as
\begin{equation}
    \mathrm{Var}\left(F\right) = \mathds{E}[F^2] - \mathds{E}[F]^2 \leq \mathds{E}[F^2] \leq 1,
\end{equation}
where in the last inequality we used that $F^2(\bm{t}) \leq 1$ as
\begin{equation}
    \int    G(\bm{t}) F^2(\bm{t}) \,  \mathrm{d}\bm{t} \leq \int    G(\bm{t}) \,  \mathrm{d}\bm{t} = 1.
\end{equation}

Let $\hat{F}$ denote the MC empirical mean estimator for the numerator using $N_{\mathrm{batch}}$ samples.
With this, the variance of the estimator can be expressed as
\begin{equation}
    \mathrm{Var}\left(\hat{F}\right) = \frac{\mathrm{Var}\left(F\right)}{N_{\mathrm{batch}}} \leq \frac{1}{N_{\mathrm{batch}}}.
\end{equation}
 A similar bound can be proven for the denominator, which is contained as a special case.

We can repeat the empirical MC estimation $K$ times for both $F$ and $J$ to obtain $K$ independent estimates of the numerator and denominator. In practice, this is done by obtaining $2KN_{\mathrm{batch}}$ independent samples, dividing them into $2K$ batches and computing the empirical mean for each batch. By taking the median over these means and computing their ratio, we get an estimate with rigorous performance guarantees, as detailed below.

\begin{figure}
	\centering
	\includegraphics[width=.4\textwidth]{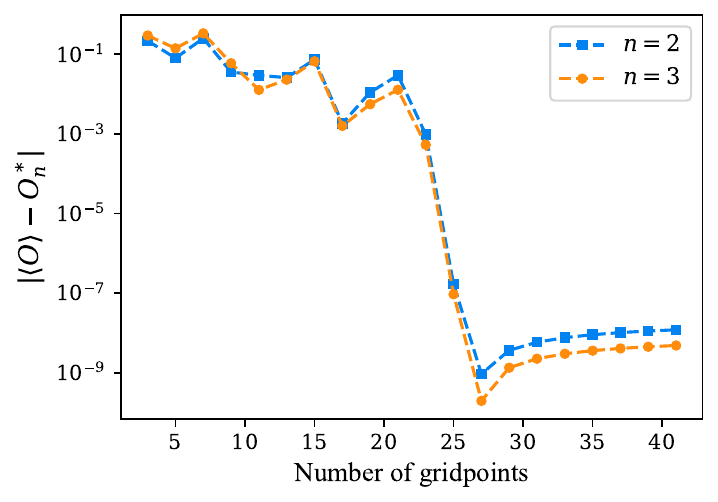}
	\caption{Deviation from the exactly achievable expectation values $O^*_n$ for $n=2$ and $n=3$ copies. The integrals for estimating the expectation value $\langle O \rangle$ are computed using the simple Riemann rule.}
	\label{fig:trapezoidal}
\end{figure}

With failure probability $\delta = e^{-K/8}$, the median of means is within $\xi$ distance from the ideal mean, where 
\begin{equation} \label{eq:xi}
\xi = 2 \sqrt{\mathrm{Var}\left(\hat{F}\right)}  \leq 2
 {N^{-1/2}_{\mathrm{batch}}} .
\end{equation}
We thus have a median of means estimate for both $F$ and $J$ with a guarantee $(\xi, \delta)$. Equivalently, for a fixed failure probability $\delta$ and precision $\xi$, we get the required number of batches $K = -8\log{\delta}$ and batch size $N_{\mathrm{batch}} = 4/\xi^2$.

In the following, let us use the notation $f = \mathds{E}[F]$ and $j = \mathds{E}[J]$. 
As such, if we now take the ratio of $F$ and $J$ with high probability, we guarantee that the error will be
\begin{align}
    \epsilon & \leq \left|\frac{f + \xi}{j - \xi} - \frac{f}{j} \right| = \sum_{i=1}^{\infty} \frac{f+j}{j^{i+1}} \xi^i, \phantom{x}\mathrm{if }  \phantom{x} f \geq 0, \\
    \epsilon & \leq \left|\frac{f - \xi}{j - \xi} - \frac{f}{j} \right| = \sum_{i=1}^{\infty} \frac{-f+j}{j^{i+1}} \xi^i, \phantom{x} \mathrm{if } \phantom{x} f < 0. \nonumber
\end{align}
Here we used the Taylor expansion of the ratio in the worst case. Since $j$ is always non-negative, only these two cases need to be considered, which can be written in a single bound as
\begin{equation}
    \epsilon \leq \left|\frac{|f| + \xi}{j - \xi} - \frac{|f|}{j} \right| = \sum_{i=1}^{\infty} \frac{|f|+j}{j^{i+1}} \xi^i .
\end{equation}

We can now substitute $\xi$ from \cref{eq:xi} and obtain a first-order estimate as
\begin{equation}
    \epsilon \leq  2 \frac{|f|+j}{j^2}
 {N^{-1/2}_{\mathrm{batch}}} + \mathcal{O}(
 {N^{-1}_{\mathrm{batch}}}).
\label{eq:median_precision}
\end{equation}
Thus, the ratio of the two median estimators has guarantee $(\epsilon, \delta)$. Inverting \cref{eq:median_precision} in the limit of large $N_{\mathrm{batch}}$, as practically relevant in classical post-processing, we obtain
\begin{equation}
    N_{\mathrm{batch}} \lessapprox 4\frac{(|f| + j)^2}{j^4} \epsilon^{-2}.
\end{equation}
This gives an estimate that is $\epsilon$-close to the ideal ratio with high probability, i.e., 
\begin{equation}
    P(|\hat{\mu}_{K,b}(F)/\hat{\mu}_{K,b}(J) - \mu(F/J)| \leq \epsilon) \geq 1-\delta.
\end{equation}

The batch sample size $N_{\mathrm{batch}}$ depends on
the expected value $|\mathds{E}[F]| =  |\tr[\ravg^n O]|$ and can be minimized in the special case where
$ \tr[\ravg^n O] \rightarrow 0$. In this vanishing limit, the bound simplifies to 
\begin{equation}
N_{\mathrm{batch}} \lessapprox 4 \tr[\ravg^n]^{-2} \epsilon^{-2}.
\end{equation}
In this case, we get the same sampling overhead, as in the \cref{lemma:complexity_vd}, therefore, in leading terms, this sample complexity can be written as $N_{\mathrm{batch}} = \mathcal{O}\left(p_q^{-2\ln{(|Q|^{-1})}} \epsilon^{-2}\right)$

\end{proof}

\begin{figure*}[t]
	\centering
	\includegraphics[width=0.9\textwidth]{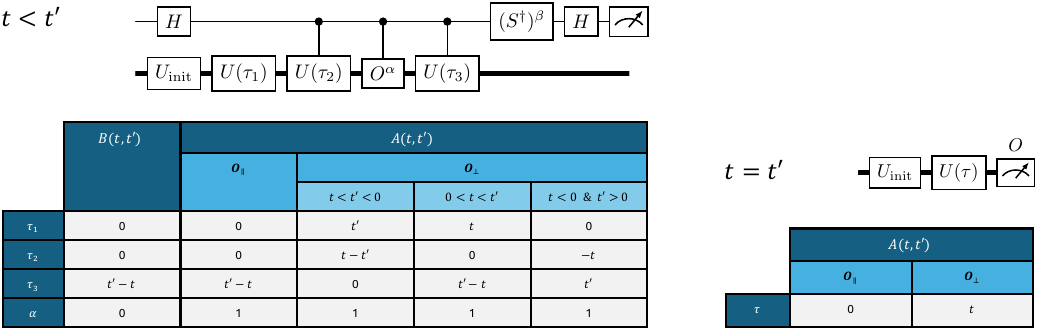}
	\caption{Quantum circuits involved in the optimized quantum computation step, each accompanied with a table of parameter values for different cases. (left) A Hadamard test used to obtain all non-trivial $A(t,t')$ and $B(t,t')$ objects with $t < t'$. The expectation value of the measurement on the ancilla qubit gives the real (imaginary) part of the chosen two-time correlator when the $S^{\dagger}$ gate is applied (not applied), corresponding to $\beta = 1$ ($\beta = 0$). The observable $O$ can be applied by decomposing it into a term that commutes ($O_{||}$) or does not commute ($O_{\perp}$) with the Hamiltonian. (right) A direct time evolution followed by a measurement of $O$ for the remaining case of $t = t'$ for $A(t,t')$. In practice, the time-evolved state can be measured independently in all of the Pauli-basis terms of $O$ and classically combined. The duration $\tau = t$ for $O_{\perp}$, but no time evolution is needed (i.e., $\tau = 0$ for all $t$) for $O_{||}$.}
	\label{fig: QC}
\end{figure*}

\section{Quadrature integration}
\label{appendix:quadrature}
Quadrature formulas can be used to approximate the integral. While it may seem natural to use Gauss-Hermite quadrature, that provides an exact formula to approximate integrals over the whole real line, it is not reliable in practice \cite{trefethen2022exactness}. Instead, we can truncate the limits to some sufficiently high number and apply Gauss–Legendre, Clenshaw–Curtis, or Riemann quadrature formulas on the truncated interval. 
With these quadrature formulas, the multidimensional integral in \cref{lemma:esd} is approximated as

\begin{align}
    \tr[ \ravg^n O]
		& =\int\limits_{-\infty}^{\infty}    G(\bm{t}) F(\bm{t}) \,  \mathrm{d}\bm{t} \\
  & \approx \int\limits_{-T}^{T}    G(\bm{t}) F(\bm{t}) \,  \mathrm{d}\bm{t} \nonumber \\
  & \approx \sum\limits_{t_1\dots t_n} A(t_n,t_1) \prod_{k=1}^{n-1}  w_{t_k}^2 G(t_k)^2 B(t_k,t_{k+1}), \nonumber
\end{align}
where the sum is taken over all combinations of time values $t_k$. Therefore, this computation is exponential in the number of copies, besides potentially introducing numerical instabilities.

In \cref{fig:trapezoidal}, we show the scaling of the error in the number of gridpoints available, using the simple Riemann rule. Here, the error is defined as the deviation from the analytically achievable estimate with the given number of copies $O^*_n$. For this demonstration, we consider the same $10$-qubit random-field Heisenberg model as in \cref{sec:exactsim}, and estimate the observable $O=Z_1$. The integrals for the numerator and denominator are approximated separately using the Riemann rule, and the estimate $\langle O\rangle$ is computed by taking their ratio.

\section{Optimized quantum processing in the hybrid quantum-classical algorithm}
\label{appendix:improvements}

An unoptimized method to obtain the $A$ and $B$ objects over a 2D grid was presented in \cref{subsection: obtaining}, but characteristics of the $A$ and $B$ objects enable significant reductions in the computational resources, including the total number of distinct quantum circuit evaluations and the durations of time evolutions involved in each evaluation. Such improvements lead to a lower susceptibility to algorithmic and hardware errors, and we demonstrate this by considering early fault-tolerant devices in \cref{subsection: gate noise}. In the following, we describe how such improvements can be made, and summarize the resulting circuits in \cref{fig: QC}. Note that, while ancilla-free implementations can be applied for problems with symmetries \cite{xu2023quantum,obrien2021error,cortes2022quantum,lu2021algorithms}, including spin models with particle-number conservation considered in this work, here we focus on a general scenario and optimize the Hadamard test.

\begin{figure}[tb]
	\centering
	\includegraphics[width=.17\textwidth]{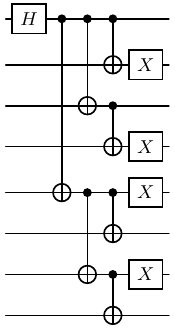}
	\caption{An example of a quantum circuit that prepares the initial state $\frac{1}{\sqrt{2}}(\ket{10100101}+\ket{01011010})$ when applied onto the all zero state $\ket{00000000}$ (i.e., $U_{\text{init}}$) considered in the Fermi-Hubbard model. The circuit can be viewed as applying bit flips onto a GHZ state.}
	\label{fig: initial_state}
\end{figure}

First, consider obtaining $B(t,t') = \bra{\psi(t)} \ket{\psi(t')}$. Since $B(t,t) = 1$ $\forall t$, we do not need to compute the diagonal of the matrix and can assume that $t\neq t'$. Furthermore, we only need to compute the case where $t' > t$ (i.e., the lower triangular part), since the other half can be obtained trivially by taking the complex conjugate as $B(t',t) = B(t,t')^*$. Finally, $\bra{\psi(t)} \ket{\psi(t')} = \bra{\psi(0)} U(t'-t) \ket{\psi(0)}$ implies that we only need to evaluate $B(t,t')$ once for each fixed $t'-t$, and the total duration associated with the Hadamard test is shorter than the unoptimized construction unless $t' > 0$ and $t < 0$, in which case they are both equal to $|t|+|t'|$. Utilizing these symmetries, only $N_T-1$ entries are sufficient to resolve the whole $B$ matrix.

Next, consider obtaining $A(t,t') = \bra{\psi(t)} O\ket{\psi(t')}$. As in the case of $B(t,t')$, we only need to obtain the lower half triangle since $A(t',t) = A(t,t')^*$. The diagonal, $A(t,t)$, can be obtained without introducing ancilla qubits, simply by applying $U(t)U_{\text{init}}$ to $\ket{0}$ and measuring the observable as shown in \cref{fig: QC} (right). To obtain the remaining entries, suppose we know a decomposition of $O$ as $O = O_{||} + O_{\perp}$, where $[O_{||},\mathcal{H}] = 0$ and $[O_{\perp},\mathcal{H}] \neq 0$. The evaluation of $A_{||}(t,t') = \bra{\psi(t)} O_{||}\ket{\psi(t')}$ can be simplified further compared to $A_{\perp}(t,t') = \bra{\psi(t)} O_{\perp}\ket{\psi(t')}$, and the desired quantity can be obtained by combining them as $A(t,t') = A_{||}(t,t') + A_{\perp}(t,t')$ for all $t' > t$. Note that $O = O_{||}$ in the important case where $[O,\mathcal{H}]=0$ (i.e., when considering symmetries). As before, each of $O_{||}$ and $O_{\perp}$ can be treated, e.g., by decomposing it into its Pauli basis or by block encoding.

Due to commutativity, $A_{||}(t,t') = \bra{\psi(0)} O_{||} U(t'-t) \ket{\psi(0)}$, so the evaluation of $A_{||}(t,t')$ is similar to that of $B(t,t')$. Particularly, the same argument as for $B$ implies that only $N_T$ entries need to be evaluated. The one additional entry needed in this case corresponds to a diagonal entry, which can be obtained from the ancilla-free circuit shown in \cref{fig: QC} (right). Moreover, this ancilla-free circuit does not require any time evolution, since commutativity implies that $A_{||}(t,t) = \bra{\psi(0)} O_{||} \ket{\psi(0)}$. For the remaining case of $t' > t$, the quantum circuit used is similar to that of $B$, but includes the additional controlled $O_{||}$ operation.

\begin{figure}[htbp]
	\centering
	\includegraphics[width=.4\textwidth]{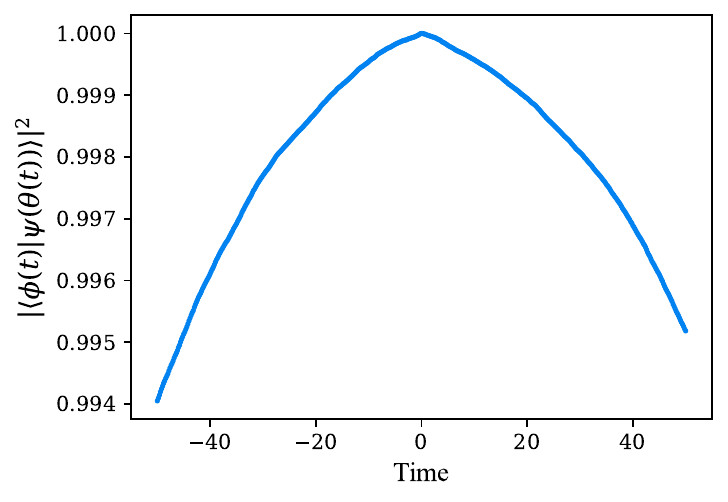}
	\caption{Fidelity of the variationally time-evolved states.}
	\label{fig:vqe_fidelity}
\end{figure}

The same argument does not hold for $A_{\perp}(t,t')$, and all $\frac{1}{2}N_T(N_T+1)$ entries of the matrix must be obtained. However, the total duration of the controlled time evolutions can still be reduced when $t$ and $t'$ have the same sign. When $0 < t < t'$, for example, $A_{\perp}(t,t') = \bra{\psi(t)} O_{\perp} U(t'-t) \ket{\psi(t)}$, such that a time evolution of duration $t$ can be performed without being controlled by the ancilla qubit, and the controlled time evolution has a duration given by the time difference $t'-t$. The same holds when $t < t' < 0$, such that the total duration of the time evolutions in each of these cases is $\text{max}\{|t|,|t'|\}$ as opposed to $|t|+|t'|$ required in the unoptimized construction. All of the simplifications discussed so far do not hold for $A_{\perp}(t,t')$ with $t < 0$ and $t' > 0$, and we apply the same circuit as the unoptimized construction just for this remaining case.

\section{Details of numerical experiments}
In all numerical experiments, the standard deviation of the normal distribution is set relative to the maximum evolution time as $\sigma = T/4$. 
We note that while the median of means estimation provides rigorous performance guarantees, it introduces an additional hyperparameter. Therefore, in the present numerical experiments we simply use an empirical mean estimator which has a similar performance in practice.
In all numerical experiments we report the performance in terms of the mean absolute error $\Delta\langle O\rangle = (1/K) \sum_i |\langle O\rangle_i - \langle O\rangle^*|$ over $K$ independent runs, where $\langle O\rangle_i$ is the $i$th independent estimate of the ratio, and $\langle O\rangle^*$ is the exact expectation value (in the case of MPS simulation it is not exact, but obtained using alternative techniques).

\begin{figure}[htbp]
	\centering
	\includegraphics[width=.45\textwidth]{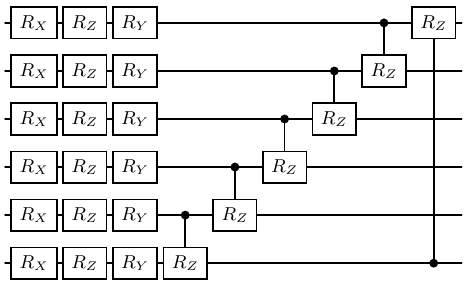}
	\caption{A single layer of the hardware-efficient ansatz used for variational quantum simulation.}
	\label{fig:he-ansatz}
\end{figure}

\subsection{Random-field Heisenberg model}
\label{appendix:spinchain}

We consider the 1D Heisenberg chain with periodic boundary conditions, random local magnetic fields and nearest neighbor interactions as
\begin{equation}
 \mathcal{H} = J \sum_{j=1}^{N} \Vec{S}_j \cdot \Vec{S}_{j+1} + \sum_j^N h_j Z_j.
\end{equation}
Here $\Vec{S}_j = \left[ X_j, Y_j, Z_j \right]$ is a Pauli vector on qubit $j$, $J$ is a coupling constant and $h_j \in [-h, h]$ is sampled uniformly at random. In our demonstrations, we use $J=0.1$ and $h=1$.

\begin{figure*}[t]
	\centering
	\includegraphics[width=\textwidth]{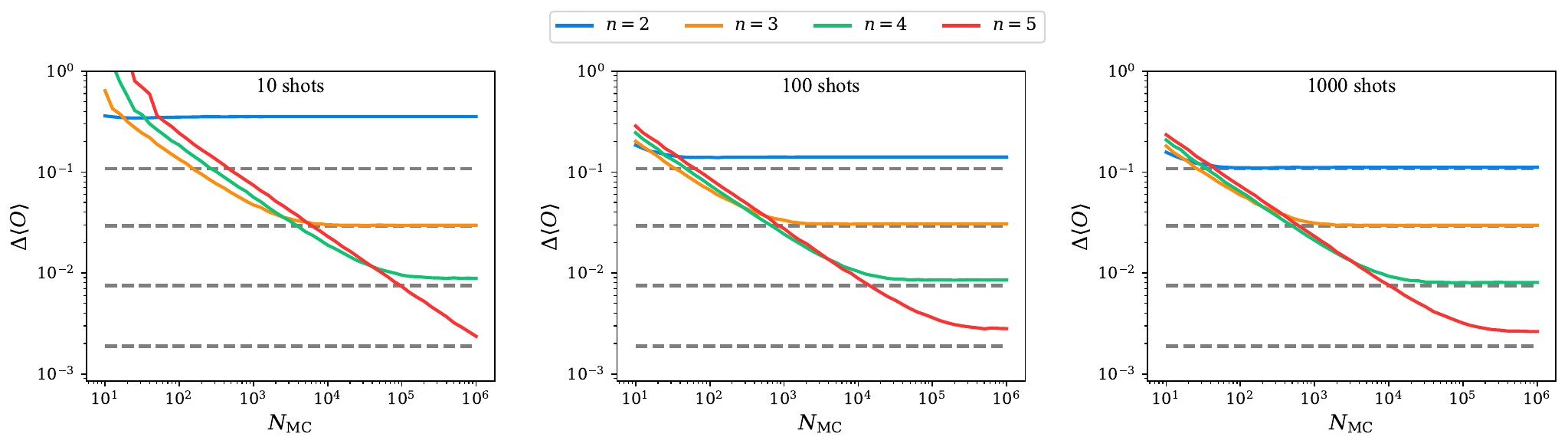}
	\caption{Exact simulation results for a $10$-qubit random-field Heisenberg model, considering shot noise. Here, we used the same settings as in \cref{fig: exact_spinring}. The effect of having a finite number of shots $N_{\mathrm{s}}$ was modeled by adding random noise sampled from a Gaussian distribution with standard deviation set to $1/\sqrt{N_{\mathrm{s}}}$.}
	\label{fig: shotnoise}
\end{figure*}

\subsection{Lattice Schwinger model}
\label{appendix:hep}
For the demonstrations involving variational methods, we considered a simple toy model in high-energy physics.
We use the lattice Schwinger model \cite{hamer1997series} in its spin Hamiltonian formulation \cite{chakraborty2022classicallyemulateddigitalquantum} 
$\mathcal{H} = \mathcal{H}_{ZZ} + \mathcal{H}_{\pm} + \mathcal{H}_{Z}$.
These individual Hamiltonian terms are defined as

\begin{align}
\mathcal{H}_{ZZ} &= \frac{J}{2} \sum_{n=2}^{N-1} \sum_{1\leq k < l \leq n} Z_k Z_l \\
\mathcal{H}_{\pm} &= \frac{J}{2} \sum_{n=1}^{N-1} (w-(-1)^n\frac{m}{2} \sin (\theta))[X_nX_{n+1} + Y_nY_{n+1}] \nonumber \\
\mathcal{H}_{Z} &= \frac{m \cos\theta}{2} \sum_{n=1}^{N} (-1)^nZ_n-\frac{J}{2}\sum_{n=1}^{N-1} (n\;\mathrm{mod}\;2)\sum_{l=1}^{n}Z_l. \nonumber
\end{align}

We set the hyperparameters heuristically to $J=1.0$, $m=0.1$, $\theta = 0.5$, and $w=0.1$.

\subsection{Fermi-Hubbard model}
\label{appendix:hubbard}
The Fermi-Hubbard model describes the behavior of interacting fermions, and is defined by the second-quantized Hamiltonian
\begin{equation}
    \tilde{\mathcal{H}}=-t\sum_{\langle i,j \rangle,s} (c^\dagger_{i,s}c_{j,s}+c^\dagger_{j,s}c_{i,s})+U\sum_i n_{i\uparrow}n_{i\downarrow}.
\end{equation}
The operators $c^\dagger_{j,s}$, $c_{j,s}$, and $n_{js} = c^\dagger_{j,s}c_{j,s}$ denote creation, annihilation, and particle-number operators, respectively, at site $j$ and spin $s$. The sum over $\langle i,j \rangle$ runs over all neighbouring lattice sites $i$ and $j$. The first term corresponds to the tunneling of fermions with tunneling amplitude $t$, and the second term accounts for on-site interactions with interaction strength $U$.

In this work, we consider a 2 by 2 square lattice with $t = 1$ and $U = 12$, and impose periodic boundary conditions. After applying the Jordan-Wigner transformation, we obtain a Hamiltonian defined on 8 qubits:
\begin{align}
\label{eq:fh_terms}
    \tilde{\mathcal{H}} =& 12 - 0.5 X_1 X_2 - 0.5 X_2 X_3 - 0.5 X_3 X_4 - 0.5 X_5 X_6\nonumber\\
    &- 0.5 X_6 X_7 - 0.5 X_7 X_8 - 0.5 Y_1 Y_2 - 0.5 Y_2 Y_3 - 0.5 Y_3 Y_4\nonumber\\
    &- 0.5 Y_5 Y_6 - 0.5 Y_6 Y_7 - 0.5 Y_7 Y_8 - 3 Z_1 - 3 Z_2\nonumber\\
    &- 3 Z_3 - 0.5 X_1 Z_2 Z_3 X_4 - 0.5 Y_1 Z_2 Z_3 Y_4 - 3 Z_4 - 3 Z_5\nonumber\\
    &+ 3 Z_1 Z_5 - 3 Z_6 + 3 Z_2 Z_6 - 3 Z_7 + 3 Z_3 Z_7\nonumber\\
    &- 0.5 X_5 Z_6 Z_7 X_8 - 0.5 Y_5 Z_6 Z_7 Y_8 - 3 Z_8 + 3 Z_4 Z_8.
\end{align}

In the time evolution of the hybrid quantum-classical algorithm, it is important to choose an initial state that has a high overlap with the target eigenstate. By choosing the lowest energy state within the subspace of four spins as the target state, we explicitly diagonalize the Hamiltonian and identify the computational basis states that have the highest overlap with it. Since two computational basis states have equally dominant overlap with the target eigenstate, we choose the initial state to be an equal superposition between them, which is $\frac{1}{\sqrt{2}}(\ket{10100101}+\ket{01011010})$ with an overlap of $p_q \approx 0.62$. Note that this is a state that can be easily prepared by applying bit flips onto a Greenberger–Horne–Zeilinger (GHZ) state, as shown in \cref{fig: initial_state}.

\subsection{Ansatz-based variational quantum simulation}
\label{appendix:varsim}

\begin{figure*}[t]
	\centering
	\includegraphics[width=0.7\textwidth]{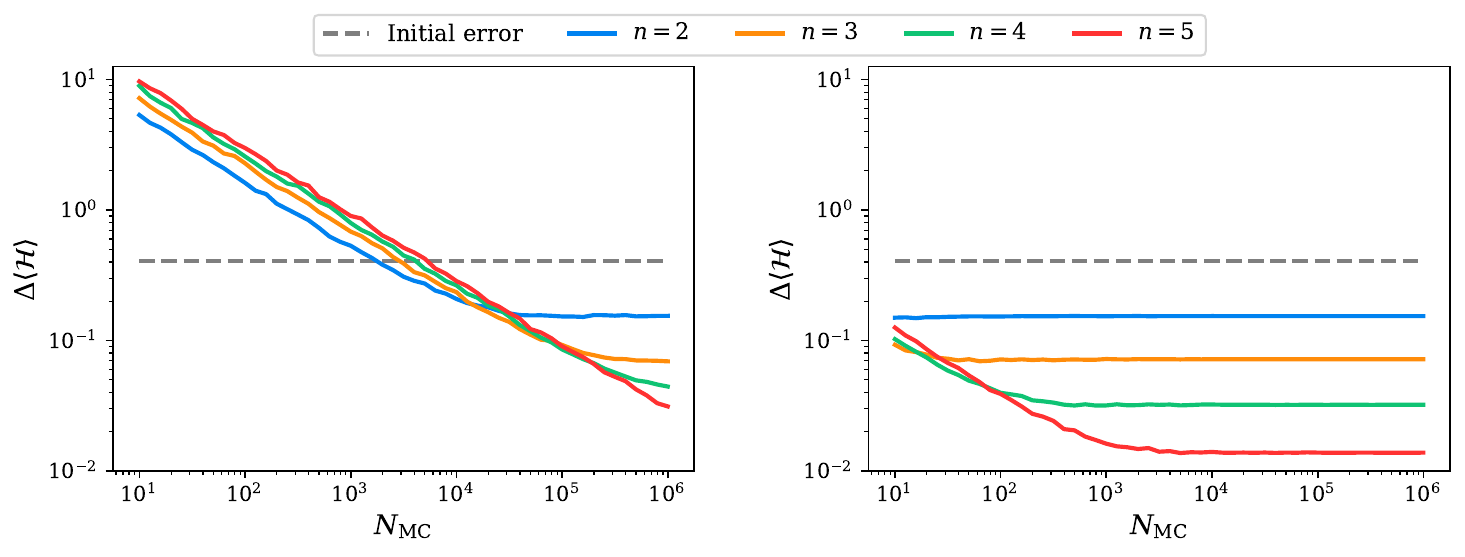}
	\caption{MPS simulation results without (left) and with (right) the variance reduction technique for the ground state of a $100$-qubit random field Heisenberg model.}
	\label{fig:mps_corr}
\end{figure*}
Variational quantum simulation leverages the variational principle to simulate the dynamics of quantum systems. This method is particularly useful for near-term quantum hardware, where implementing deep quantum circuits may not be feasible.

In our demonstration, we rely on an ansatz-based approach, using the hardware-efficient ansatz with ring topology, as shown in \cref{fig:he-ansatz}. Using the time grid with $T=50$ and $dt = 0.2$, we define the step size of the simulation algorithm as $\delta t = dt/1000$. Having the optimized parameters at $t=0$, we seek to estimate the parameters corresponding to the time-evolved state at each step.
For this, we need to solve, the linear system of equations
\begin{equation}
    \sum_{j} A_{i,j}^{\mathrm{R}} \dot{\theta}_j = C_i^{\mathrm{I}} ,
    \label{eq:varsim}
\end{equation}
where
\begin{equation}
    A_{i,j} = \frac{\partial \bra{\psi(\boldsymbol{\theta}(t))}}{\partial \theta_i} \frac{\partial \ket{\psi(\boldsymbol{\theta}(t))}}{\partial \theta_i}
\end{equation}
is the quantum metric tensor, and 
\begin{equation}
    C_{i} = \frac{\partial\bra{\psi(\boldsymbol{\theta})}}{\partial \theta_i} H \ket{\psi(\boldsymbol{\theta})}.
\end{equation}
is the gradient vector. Here $A_{i,j}^{\mathrm{R}}$ and $C_i^{\mathrm{I}}$ denote the real and imaginary parts respectively. By solving for the time derivatives $\dot{\boldsymbol{\theta}}$, we can compute the updated parameters at the next timestep as $\theta'_i = \theta_i+\dot{\theta}_i \delta t$. In practical scenarios, the quantum metric tensor can be ill conditioned, therefore instead of solving \eqref{eq:varsim} we find the time derivatives, that minimize the Tikhonov-regularized system
\begin{equation}
    \sum_{j} M_{i,j}^{\mathrm{R}} \dot{\theta}_j = C_i^{\mathrm{I}} ,
\end{equation}
where $M = A + \lambda \mathds{1} $, and we set $\lambda=10^{-4}$.

It is obvious from this update rule that the fidelity of the states in this iterative process highly depends on the step size. Therefore, we chose $\delta t$ small enough to maintain high fidelity. The fidelity of the variationally time-evolved states $\ket{\psi(\boldsymbol{\theta}(t))}$ compared to the exact time evolution $\ket{\phi(t)}$ is shown in \cref{fig:vqe_fidelity}.

\subsection{Early fault-tolerant quantum devices}
\label{appendix:early}

For our simulations in \cref{subsection: gate noise}, we assume that the dominant error source of an early fault-tolerant device is the approximation of continuous rotation gates by sequences of Clifford+T gates, which noise model is often assumed in the literature~\cite{PRXQuantum.5.040352,PRXQuantum.6.010352,goh2024direct,kliuchnikov2023shorter}. Since the magic state distillation required to implement T gates with sufficiently suppressed errors is costly, an early fault-tolerant device would have to resort to shallow approximation sequences. This results in deviations from the ideal continuous rotation gate, which we model by applying small depolarizing errors after continuous rotation gates~\cite{cai2020mitigating,kliuchnikov2023shorter,PRXQuantum.5.040352} are applied.

Both the unoptimized construction described in \cref{subsection: obtaining} (shown in the quantum computation step of \cref{fig:algo}) and the optimized construction given in Appendix~\ref{appendix:improvements} (shown in \cref{fig: QC}) involve either direct or controlled time evolutions generated by the Hamiltonian with the term proportional to the identity operator removed. When applying Trotterization as in \cref{subsection: trotter error} for the Fermi-Hubbard Hamiltonian in \cref{eq:fh_terms}, each continuous rotation gate is generated by a Pauli string of the form $S_iZ_{i+1}\ldots Z_{j-1}S_j$ with $S \in \{X,Y,Z\}$. Errors for rotations that correspond to terms with $i=j$ or $j = i + 1$ are modeled by single-qubit depolarizing errors occurring with equal probabilities on qubits $i$ and $j$ (without double-counting when $i = j$). Since a network of fermionic $\textsc{SWAP}$ gates are able to remove the $Z$ operators sandwiched between $S_i$ and $S_j$ in the rest of the terms \cite{verstraete2009quantum,kivilichan2018quantum,cai2020resource}, we model errors for these more general rotation terms similarly by only applying depolarizing errors on qubits $i$ and $j$. When the time evolution is controlled by the state of an ancilla qubit, the ancilla qubit is also subject to a depolarizing error with the same error probability.

\section{Additional numerical results}
\label{appendix:numerics}
\subsection{Exact simulation with shot noise}

We also consider shot noise arising from the finite number of quantum circuit evaluations for the random-field Heisenberg model, re-running the MC estimator with noisy data.
For a single Pauli string, sampling the quantum state corresponds to sampling from a binomial distribution $\mathcal{B}(N_{\mathrm{s}}, p)$, and this can be approximated by a Gaussian distribution $\mathcal{N}(N_{\mathrm{s}}p, N_{\mathrm{s}}p(1-p))$ for large number of shots $N_{\mathrm{s}}$. Considering the upper bound $p(1-p)<1$, we model the effect of shot noise on the corresponding $A$ and $B$ objects by sampling from the Gaussian distribution with standard deviation $1/\sqrt{N_{\mathrm{s}}}$.
We consider cases where both the real and imaginary parts of each entry of the $A$ and $B$ correlators are estimated using $10$, $100$ or $1000$ uncorrelated shots. The corresponding results are shown in \cref{fig: shotnoise}. 

With a standard deviation corresponding to only $1000$ shots, our MC method shows similar performance to the noiseless case. Considering fewer shots, we get irregular behavior that does not necessarily reach the theoretical performance of the noiseless results, but in most cases, it still suppresses the initial error along with the shot noise itself. 

\subsection{MPS simulations with variance reduction}
\label{appendix:mps_numerics}
We compare the effect of the variance reduction method described in \cref{sec:variance} for the ground state estimation with MPS simulation. In \cref{fig:mps_corr} we show the error compared to the DMRG energy without (left) and with (right) the variance reduction for different number of copies and increasing number of MC samples. It is clear, that the initial bias without the reduction technique can be substantial, when it is not yet suppressed by the number of samples.


%

\end{document}